\documentstyle[emulateapj]{article}

\def\lesssim{\mathrel{\hbox{\rlap{\hbox{\lower4pt\hbox{$\sim$}}}\hbox{$<$}}}}
\def\gtrsim{\mathrel{\hbox{\rlap{\hbox{\lower4pt\hbox{$\sim$}}}\hbox{$>$}}}}

\slugcomment{submitted to The Astrophysical Journal; Draft version \today}

\begin{document}

\title{
On the Initial Mass Function of Population III Stars
}
\author{Fumitaka Nakamura}
\affil{Faculty of Education and Human Sciences, Niigata University,
8050 Ikarashi-2, Niigata 950-2181, Japan}
\and
\author{Masayuki Umemura}
\affil{Center for Computational Physics, University of Tsukuba,
Tsukuba, Ibaraki-ken 305-8577, Japan}

\begin{abstract}

The collapse and fragmentation of filamentary 
 primordial gas clouds are explored using one-dimensional and 
 two-dimensional hydrodynamical simulations coupled with the
 nonequilibrium processes of hydrogen molecule formation.
The cloud evolution is computed from the initial central density 
 $n_c=10 - 10^6$ cm$^{-3}$.
The simulations show that depending upon the initial density,
there are two occasions for the fragmentation
of primordial filaments. If a filament has relatively low initial density
such as $n_c \lesssim 10^5$ cm$^{-3}$, the radial contraction 
is slow due to less effective H$_2$ cooling and appreciably 
decelerates at densities higher than a critical density,
 where LTE populations are achieved
 for the rotational levels of H$_2$ molecules and 
 the cooling timescale becomes  accordingly longer
 than the free-fall timescale. 
This filament tends to fragment into dense clumps
 before the central density reaches $10^{8-9}$ cm$^{-3}$,
 where H$_2$ cooling by three-body reactions is effective 
 and the fragment mass is more massive than some tens $M_\odot$. 
In contrast, if a filament is initially as dense
as $n_c \gtrsim 10^5$ cm$^{-3}$, the more effective H$_2$ cooling 
with the help of three-body reactions allows the filament to contract 
up to $n\sim 10^{12}$ cm$^{-3}$. After the density reaches
 $n\sim 10^{12}$ cm$^{-3}$, the filament becomes optically thick
 to H$_2$ lines and the radial contraction subsequently almost stops.
At this final hydrostatic stage, 
the fragment mass is lowered down to $\approx 1M_\odot$
because of the high density of the filament. 
The dependence of the fragment mass upon
the initial density could be translated into the dependence on the local
amplitude of random Gaussian density fields or the epoch of the collapse of
a parent cloud.
Hence, it is predicted that the initial mass function of
 Population III stars is 
likely to be bimodal with peaks of $\approx 10^2 M_\odot$
and $\approx 1M_\odot$, where the relative heights 
could be a function of the collapse epoch.
Implications for the metal enrichment by Population III stars 
at high redshifts and 
baryonic dark matter are briefly discussed.

\end{abstract}

\keywords{cosmology: theory --- galaxies: formation 
--- hydrodynamics --- ISM: clouds --- stars: formation}

\section{Introduction}

The existence of a very first generation of objects, namely Population III, 
has been originally
postulated by the presence of a noticeable amount of heavy elements
in Population II stars (see, e.g., Carr, Bond, \& Arnett 1984 and
references therein) 
and has recently gained increasing importance
owing to the discovery of intergalactic metals in the Ly$\alpha$ forest
(Cowie et al. 1995; Songaila \& Cowie 1996). 
In those respects, Population III objects need to be such 
 massive stars that they can produce metals at the end of their evolution. 
Recent theoretical analyses on the evolution of metal-free stars predict that 
the fate of the massive metal-free stars can be classified as follows
 (e.g., Heger, Woosley, \& Waters 2000; Chiosi 2000; see also 
 Portinari, Chiosi, \& Bressan 1998 for the effects of mass loss.): 
(1) A star with a mass of $m \gtrsim 250 M_\odot$ 
 collapses completely to a black hole (BH) without 
 ejecting any heavy elements (Bond, Carr \& Arnett 1984).
(2) A star of $100M_\odot \lesssim m \lesssim 250 M_\odot$
 is partly or completely disrupted by 
 electron-positron pair instability.
For $m \gtrsim 150 M_\odot$, the core completely disrupts,
  and the whole core involving heavy elements
is injected in the intergalactic medium.
If it is an extremely energetic event (a hypernova),
it might lead to a gamma ray burst (GRB).
(3) A star of $35M_\odot \lesssim m \lesssim 100 M_\odot$ 
 probably collapses into a black hole.
(4) A star of $10M_\odot \lesssim m \lesssim 35 M_\odot$ results
 in a type II supernova. 

Population III stars are related to various issues that are
currently the object of considerable attention.
The luminous Population III stars could cause the reionization 
of the universe at redshifts $\gtrsim 5$ 
(Couchman \& Rees 1986; Fukugita \& Kawasaki 1994; 
Ostriker \& Gnedin 1996; Gnedin \& Ostriker 1997;
Haiman \& Loeb 1997; Miralda-Escud\'{e} \& Rees 1998; Gnedin 2000).
Alternatively, moderately massive BHs as the end products of massive 
stars might coagulate into a super-massive BH, evolving to 
primordial AGNs (Larson 2000). The accreting super-massive BHs 
may be more responsible for cosmic reionization 
(Tegmark \& Silk 1994; Sasaki \& Umemura 1996; Haiman \& Loeb 1998;
Valageas \& Silk 1999). 
In addition, Population III stars may play an important role 
 in the early evolution of galaxies 
 (e.g., Tegmark, Silk, \& Blanchard 1994; Ostriker \& Gnedin 1996)
 or the early formation of massive BHs of $\sim 10^5 M_\odot$
 (Umemura, Loeb, \& Turner 1993).
They may be responsible for the observed abundance patterns
 of extremely metal-deficient stars
 (McWilliam et al. 1995; Ryan, Norris, \& Beers 1996; 
 Audouze \& Silk 1995; Shigeyama \& Tsujimoto 1998).
Finally, if a significant number of MACHOs are ancient white dwarfs
(M\'{e}ndez \& Minniti 2000), they may stem from low-mass 
Population III stars (Carr 1994; Larson 1998; Rees 1999). 
In the light of all of such possible
significant consequences,
the initial mass function (IMF) of Population III stars
is an issue of undoubted importance.

Many authors have studied the collapse
 of primordial clouds to estimate the masses of Population III stars
 (e.g., Matsuda, Sato, \& Takeda 1969; Yoneyama 1972;
 Hutchins 1976; Silk 1977, 1983; Carlberg 1981; 
 Palla, Salpeter, \& Stahler 1983;
 Yoshii \& Saio 1986; Nishi et al. 1998).
These studies have emphasized the importance of
 radiative cooling by H$_2$ molecules because the primordial gas 
 is deficient in heavy elements, which are the most efficient
 coolants in present-day star formation.
In the first collapsed objects, baryonic gas is heated
 to temperatures above $10^3$ K during the contraction.
This enhances the H$_2$ formation rate
 and causes the H$_2$ abundance 
 to rise from its initial value of $\sim 10^{-6}$ to 
 a quasi-equilibrium value of $10^{-4}\sim 10^{-3}$
(e.g., Palla et al. 1983; Lepp \& Shull 1984; Haiman, Thoul \& Loeb 1996).
Thus, H$_2$ molecules cool the gas to a temperature of 
 $\sim 500$ K.
As a result, the Jeans mass descends to a stellar mass
 and Population III stars can form in this way
 through fragmentation of the first collapsed objects.
Although many elaborate analyses have been made, 
the estimated masses of Population III stars
 have not been well converged.
Several authors have suggested that Population III stars
 were low mass, 
 whereas others have suggested that they were massive or 
 very massive.
This discrepancy seems to come from most studies being
 restricted to highly simplified models such as 
 homogeneous, pressureless, and/or spherical collapses.

In the bottom-up scenarios like cold dark matter (CDM) models,
 the first collapsed pregalactic objects should form 
 at redshifts of $z\sim 10-10^2$ and have masses of 
 $10^5 \sim 10^8M_\odot$ (Tegmark et al. 1997).
Because of the asymptotic scale-invariance of the CDM density 
fluctuations, pregalactic clouds with this mass range could
undergo a run-away collapse, forming cores of
$\sim 10^2 M_\odot$
 (Abel et al. 1998; Abel, Norman, \& Bryan 2000),
or form mini-pancakes, fragmenting into pieces of
$10^2 M_\odot-10^3 M_\odot$ (Bromm, Coppi, \& Larson 1999).
In these calculations, one of the common features is
 the formation of a filamentary structure. 
Filamentary clouds are gravitationally unstable and likely to 
 fragment into dense clumps. 
Such dense clumps are expected to evolve into Population III stars.
The physics of the fragmentation of filamentary primordial clouds 
has been studied analytically
(Uehara et al. 1996) or by a one-dimensional numerical
simulation (Nakamura \& Umemura 1999, hereafter Paper I).
These studies suggest that the minimum mass of Population III stars
is of the order of $M_\odot$. However,
the physical processes are still unclear to determine whether
Population III stars can be far above $M_\odot$ or 
be eventually reduced to a few $M_\odot$.
Thus, as a further step following Paper I, we here perform two-dimensional
hydrodynamical simulations.
Attention is focused on elucidating
the physical process of the fragmentation of
primordial gas filaments to assess the mass of Population III stars.

In Paper I, we pursued the radial contraction 
 of primordial filaments and 
showed that the filaments continue to contract
 quasi-statically, the temperatures staying nearly constant 
 at $T\sim 500$ K.
When the cloud becomes optically thick to the H$_2$ lines,
 the radial contraction essentially stops.
Applying a linear stability analysis,
 the fragmentation was expected to take place at that stage and 
 the minimum masses of Population III stars
 were estimated as a few $M_\odot$.
Thus, Population III stars were anticipated to be 
 low-mass deficient compared to the present-day stars.
In this paper, we pursue the fragmentation processes of the filaments
 by means of two-dimensional axisymmetric simulations.
The present model is an improved version of Paper I.
The numerical model and method are described in \S 2.
Numerical results are given in \S 3 and \S 4.
We show that filaments with low initial density can fragment
 into dense clumps before the cloud becomes optically
 thick to the H$_2$ lines.
Then, the masses of the clumps
could be much more massive than a few $M_\odot$.
However, relatively dense filaments result in clumps of a few $M_\odot$.
Hence, in \S 5, it is predicted that the initial mass function
 of Population III stars is likely to be bimodal. 
In \S 6, we discuss some implications for the first metal
 enrichment and baryonic dark matter.

\section{Model and Numerical Methods}
\label{sec:model}

\subsection{Basic Equations}
\label{subsec:basic}

To pursue the collapse and 
 fragmentation of filamentary primordial clouds, 
 we employ a two-dimensional hydrodynamical scheme.  
We assume that the system is axisymmetric and 
 that the medium consists of ideal gas.
The adiabatic index, $\gamma$, is taken to be 5/3 for
 monatomic gas and 7/5 for diatomic gas.
We deal with the following 9 species:
 e, H, H$^+$, H$^-$, H$_2$, H$_2^+$, He, He$^+$, and He$^{++}$.
The mass fraction of He is taken to be 0.24
 of the total mass.

The basic equations are then described as 
\begin{equation}
 \frac{\partial \rho}{\partial t}+\nabla \cdot 
  (\rho \mbox{\boldmath$v$})=0 \,  ,
  \label{eq:basic1}
\end{equation}
\begin{equation}
 \frac{\partial (\rho {\mbox{\boldmath$v$}})}{\partial t}+\nabla \cdot 
  (\rho \mbox{\boldmath$v$}\otimes \mbox{\boldmath$v$})
  +\nabla P +\rho \nabla \psi =0 \, ,
  \label{eq:basic2}
\end{equation}
\begin{equation}
 \frac{\partial E}{\partial t}+\nabla \cdot 
  \left[(E+P)\mbox{\boldmath$v$}\right]
  +\rho \mbox{\boldmath$v$} \cdot \nabla \psi +\Lambda _{\rm net}=0 \, ,
  \label{eq:basic3}
\end{equation}
\begin{equation}
 \triangle \psi =4\pi G\rho \, ,
  \label{eq:basic4}
\end{equation}
\begin{equation}
 \rho = \sum _i \rho _i \, ,
  \label{eq:basic5}
\end{equation}
\begin{equation}
 P=\sum _i P_i=\sum _i n_i k T \, ,
  \label{eq:basic6}
\end{equation} 
\begin{equation}
 E=\sum _i \left(\frac{P_i}{\gamma _i -1}+\frac{1}{2}
\rho _i \mbox{\boldmath$v$}^2 \right) \, ,
  \label{eq:basic7}
\end{equation}
 where $\rho$ is the mass density, $n$ is the number density, 
 $\mbox{\boldmath$v$}$ is the velocity, $P$ is the gas pressure, 
 $\psi$ is the gravitational potential, $E$ is the energy per unit
 volume, $G$ is the gravitational constant, and $k$ is the 
 Boltzmann constant.
The symbol $\Lambda _{\rm net} $ denotes the
 net energy-loss rate per unit volume.
The values with subscript $i$ denote those of the $i$-th species.

The number density of the $i$-th species, $n_i$, is obtained 
by solving the following time-dependent rate equations,
\begin{equation}
 \frac{dn_i}{dt}= \sum ^{9} _{j=1} \sum ^{9} _{k=1} k_{jk} n_j 
  n_k+ \sum ^{9} _{l=1} \sum ^{9} _{m=1} \sum ^{9} _{n=1}
  k_{lmn} n_l n_m n_n \, ,
\label{eq:basic8}
\end{equation}
 where the reaction rate coefficients,  $k_{jk}$ and $k_{lmn} $, 
 are given in Table \ref{tab:1}.
The relative abundances of hydrogen species 
 are given by $ x_i \equiv n_i / n $, where
$n \equiv n_{\rm H}+
 n_{\rm H^+}+n_{\rm H^-}+2n_{\rm H_2}+2n_{\rm H_2^+}$.

We take into account the following thermal processes
 by H atoms and H$_2$ molecules: 
 (1) H cooling by recombination, 
 collisional ionization, and collisional excitation (Cen 1992), 
 (2) H$_2$ line cooling by rotational and
 vibrational transitions (see below), 
 (3) cooling by H$_2$ collisional dissociation
 (Shapiro \& Kang 1987), and 
 (4) heating by H$_2$ formation (Susa, Uehara, \& Nishi 1996).
In Paper I, for the H$_2$ formation heating, we only took into 
 account the contributions by two-body reactions
 (Shapiro \& Kang 1987).
In this paper, we include the contributions by three-body 
 reactions, which play an important role
 in temperature evolution
 after the density reaches $10^9$ cm$^{-3}$ (see \S 3).
Other thermal processes are negligible 
 because the gas temperatures did not exceed 10$^4$ K
 in the models calculated in this paper.

As shown by many authors, H$_2$ line cooling
 is most efficient in primordial gas.
Therefore, careful evaluation of the H$_2$ line cooling rate
 is necessary.
We thus compute the H$_2$ line cooling rate as follows.
First, the level populations at the rotational and 
 vibrational excitations are determined by 
 using a recursion formula (Hutchins 1976; 
Palla et al. 1983). 
For the collisional deexcitation rates, we consider both H-H$_2$ and H$_2$-H$_2$ collisions 
 using the analytical fits of Hollenbach \& McKee (1979) and Galli \& Palla (1998).
The 21 rotational and 3 vibrational levels are taken into account.
Next, a photon escape probability method is applied 
 for each transition (Castor 1970; Goldreich \& Kwan 1974).
Finally, the line cooling rate is computed as
\begin{equation}
\Lambda _{\rm H_2 line} =\sum \Lambda _{ij} 
= \sum n_{l, i} A_{ij}
 h\nu _{ij} \beta _{ij} \; ,
\end{equation}
 where $n_{l, i}$ is the H$_2$ level population at level $i$,
 $A_{ij}$ is the Einstein A-coefficient,
 and $h\nu _{ij}$ is the energy difference between levels $i$ and $j$.
The photon escape probability $\beta _{ij}$ 
 is defined as 
\begin{equation}
 \beta _{ij}=\frac{1-\exp (-\tau _{ij})}
 {\tau _{ij}} \; ,
\end{equation}
and 
\begin{eqnarray}
 \tau _{ij}(r,z)&=& \int _r ^{\rm r_{\rm out}}
 \alpha _{ij}(r',z)dr' \\ 
 &=& \int _r ^{\rm r_{\rm out}}
\frac{h\nu _{ij}}{8\pi\Delta \nu _{ij}}
(n_{l,j}B_{ji}-n_{l,i}B_{ij}) dr' \; ,
\end{eqnarray}
 where $\tau _{i\rightarrow j}$ is the optical depth at the transition 
 $i\rightarrow j$, $r_{\rm out}$ is the radius of the cloud surface, 
 $\alpha _{i\rightarrow j}$ is the absorption coefficient 
 of the transition $i\rightarrow j$, $B_{ij}$ and $B_{ji}$ are
 the Einstein B-coefficients, and $\Delta \nu _{ij} (\propto T^{1/2})$ 
 is the thermal Doppler width of the transition line $i\rightarrow j$. 
Our cooling function almost coincides with that of Galli \& Palla (1998) 
 as long as the cloud is optically thin to the H$_2$ lines.
However, once the cloud becomes opaque to the cooling radiation, 
 our cooling rate significantly deviates from that of Galli \& Palla (1998).

\subsection{Model of Filamentary Primordial Clouds}
\label{subsec:filament}

In bottom-up scenarios such as CDM models, 
 the first collapsed objects at $z \sim 10 - 100 $
 are expected to have masses of $10^{5-8} M _\odot$.
In these clouds, the gas is heated above $T>10^{3-5}$ K by shock.
Therefore, just after the shock formation, the cooling timescale 
 is likely to be shorter than the free-fall one.
Then, the gas is cooled by the H$_2$ cooling down 
 to the temperature at which the cooling timescale 
 is comparable to the free-fall one
 (e.g., Haiman et al. 1996; Yoneyama 1972).
The resultant gas temperature is estimated to be 
 $T\simeq 10^{2-3}$ K, depending weakly on the gas density and 
 the H$_2$ fraction.
Recent numerical simulations have shown that such clouds 
 tend to fragment into filaments that collapse
 toward the major axes
  (e.g., Bromm et al. 1999; Tsuribe 2000).
The collapsing filaments are expected to fragment into
 denser clumps where Population III stars will form.
A model of such a filamentary cloud is described below.

The model of a filamentary cloud is basically the same as that presented in Paper I.
We consider an infinitely long cylindrical gas cloud
 that is collapsing in the radial direction. 
The initial temperature and relative abundances
 are assumed to be spatially constant.
At the initial state, the relative abundances of 
 H$^-$, H$_2^+$, He$^+$, and He$^{++}$ are set to zero 
 for simplicity.
We do not consider the effect of dark matter because after
 virialization of the parent system,
the local density of baryonic gas is likely to become higher than the
background dark matter density owing to radiative cooling 
(e.g., Cen \& Ostriker 1992a, 1992b; Umemura 1993).
The density is assumed to be uniform along the cylinder axis 
and its radial distribution is expressed as 
\begin{equation}
\rho =\rho _0 \left(1+\frac{r^2}{R_{0}^2}\right)^{-2} \, ,
\end{equation}
where $R_{0}=\sqrt{2fkT_0/(\pi G\rho _0 \mu)}$
 is the effective radius, 
 $\rho_0$ is the central mass density, 
 $T_0$ is the initial gas temperature, 
 $\mu$ is the mean molecular weight, 
 and $f$ is the ratio of the gravitational force to the pressure force.
When $f=1$, the density distribution coincides with
 that of an isothermal filament in equilibrium (Stod\'o\l kiewicz 1963).  
In this paper, we restrict the parameter $f$
 to $f> 1$ since we are interested in the evolution of 
 collapsing clouds.

The radial infall velocity is given by
\begin{equation}
v_r =-\frac{v_0 r}{R_0 +\sqrt{R_0^2+r^2}}\, ,
\end{equation}
which has a qualitatively similar distribution to
 that of a self-similar solution of a collapsing filament
 (see Figure 6 of Paper I).
Here, $v_0$ is constant and is set to 
 an initial sound speed ($c_s$) for all the models calculated 
 in this paper. 
We also calculated the evolution of the models with 
 different velocity distributions, e.g.,
 $v_r =-v_0 \sin (\pi r/R_{\rm max})$ and 
 confirmed that the numerical results
 are not sensitive to the assumed velocity profiles 
 as long as $v_0 \lesssim {\rm a \ few} \times c_s$.
This is because even if the initial cloud is static,
 the radial contraction is immediately accelerated 
 owing to the H$_2$ cooling.

For the above model, 
 the mass per unit length (line mass) is given by
\begin{eqnarray}
 l_0 & = & \int ^\infty _0 2 \pi \rho r dr 
 =  \frac{2fkT_0}{\mu G} \nonumber \\
& = & 1.8 \times 10^3  M_\odot {\rm pc^{-1}} \left(\frac{f}{2}\right)
\left(\frac{T_0}{300 {\rm K}}\right) \, .
\label{eq:line mass}
\end{eqnarray}
Note that the line mass of the equilibrium ($f=1$)
 filament depends only on the temperature.
When the filamentary cloud forms through the gravitational fragmentation of
 a parent sheetlike cloud, its line mass is expected to be
 nearly twice that of the equilibrium cloud
 (see \S 2.2 of Paper I).
We thus adopt $f\sim 1.5$ for typical models.

This model is then specified by the five parameters: $n_0\equiv\rho _0/\mu$,
 $T_0$, $f$, $x_e$, and $x_{\rm H_2}$.
The abundances of $x_e$, $x_{\rm H}$, and $x_{\rm H^+}$ are determined by 
 the conservation of mass and charge.
The initial density of the filament 
 depends sensitively on the properties
 of the parent cloud such as the mass, collapse redshift,
and spin parameter (see the Appendix).
Thus, we take a wide range of the initial density, 
 $10$ cm$^{-3}$ $\le n_0\le$ $10^6$ cm$^{-3}$.
Previous numerical simulations 
 (e.g., Haiman, Rees, \& Loeb 1996; Abel et al. 1998; 
Bromm et al. 1999)
have shown that the temperatures and H$_2$ relative abundances 
 reach $T\sim 300- 500$ K and $x_{\rm H_2}\sim 10^{-3}-10^{-4}$,
 respectively, when the first collapsed objects with masses
 of $10^{5-8}M_\odot$ are virialized.
Thus, the initial temperatures and H$_2$ abundances 
 are set to $T= 300\sim 500$ K and $x_{\rm H_2}\sim 10^{-3}-10^{-4}$,
 respectively.
The electron abundance is set to $5\times 10^{-5}$ for all models.
As mentioned in Paper I, the numerical results do not depend 
 sensitively upon $x_e$ as far as $x_e \gtrsim 10^{-7}$, 
 because free electrons can quickly recombine
 to a level of $x_e \lesssim 10^{-7}$
 in the course of the collapse.

\subsection{Density Fluctuations and Initial Parameters}

The initial model of the computation is constructed
 by imposing linear density fluctuations 
 on the model filament described in the above subsection.
The power spectrum of the density fluctuations
 is assumed to be a power law distribution of 
 $P(k)=Ak^\nu$ on a wavenumber space, where 
 $k$ is the wavenumber of the density fluctuation
 in the $z$-direction and $\nu$ is the power index.
The phase of the fluctuations has a random distribution 
 in the range of 0 to $2\pi$.
In this paper, the power index $\nu$ is taken to be $\nu=-2$, $-1$, and 0.
(Note that $\nu=-1$ provides scale-invariant fluctuations in
the present geometry.)
The amplitude is specified by 
$\delta \equiv [P(k)k]^{1/2}$ at $k=2\pi/R_0$.
The simulations were done in the range of $0.1 \geq \delta \geq 0.001$.

Table \ref{tab:model} summarizes the initial parameters
 of the numerical models calculated in this paper.

\subsection{Numerical Methods}

The hydrodynamic equations (\ref{eq:basic1}) 
through (\ref{eq:basic7}) are solved numerically 
using a second-order upwind scheme based on 
the method by Nobuta \& Hanawa (1999).  
See Nobuta \& Hanawa (1999) for more details
 and the test of the code.
The rate equations (\ref{eq:basic8}) are 
solved numerically with the LSODAR (Livemore Solver for
Ordinary Differential equations with Automatic method switching for
stiff and non-stiff problems) coded by L. Petzold and A. Hindmarsh. 

The computations are performed on a cylindrical domain 
($0\le r\le R_{\rm max}$ and $0\le z\le Z_{\rm max}$)
with a fixed boundary condition at $r=R_{\rm max}$ and 
periodic boundary conditions at $z=0$ and $Z_{\rm max}$, 
where $R_{\rm max}$ and $Z_{\rm max}$ are
 the maximum $r$- and $z$-coordinates on the computational domain, respectively.
Uniform grids are employed for all runs.
In all the models we take $R_{\rm max}$ 
to be more than three times the effective radius $R_{\rm 0}$. 
The effect of the fixed boundary is very small.
This is because the density is much lower near the outer boundary 
($r\sim R_{\rm max}$) than at the center, i.e.,
$\rho\lesssim 10^{-3} \rho _c $ for all runs.
In fact, we have calculated the cases with larger $R_{\rm max}$
and have confirmed that the numerical results are not affected.

\section{Radial Contraction of Filamentary Clouds}
\label{sec:result1}

Before examining the fragmentation processes of filamentary clouds,
we pursue the evolution of the filamentary clouds with no density
fluctuations, i.e., their radial contraction.
These calculations are the extension of Paper I.
(In Paper I, we explored the evolution of the filaments 
 in the initial density range of
 $10^2$ cm$^{-3}$ $\le n_0 \le 10^4$ cm$^{-3}$.
In this paper, we extend the initial density range to 
 $10$ cm$^{-3}$ $\le n_0 \le 10^6$ cm$^{-3}$.)
In the following we focus on the evolution of physical quantities 
on the $z$-axis. 
See Paper I for more details, e.g., the radial
 distributions of the physical quantities.
As shown below, the evolution of the filaments can be 
 classified into two types, depending mainly on the initial 
 density.
 
\subsection{A Low-Density Filament:
  A Quasi-Statically Contracting Filament (Model A1a)}

As a typical example of low-density filaments,
 we show the evolution of model A1a 
 which has the initial parameters of 
 $n_0=10$ cm$^{-3}$, $T_0=400$ K, and $f=1.5$.
The electron and H$_2$ number fractions are initially set to 
$5 \times 10^{-5}$ and $10^{-4}$, respectively.
Figure 1a shows the evolution of the temperature
 and the H$_2$ number fraction 
 as a function of the central density.
The central density monotonously increases with time.
Therefore, the abscissa corresponds to the evolution time.
For comparison, we pursued the evolution of 
 model A1a in which the three-body reactions
 are not taken into account.
The evolutional paths are shown by dotted lines
 in Figure 1.

Figure 1b shows the evolution of
 three characteristic timescales as a function of 
 the central density.
The solid, dashed, and dashed-dotted lines denote
 the contraction time, the cooling time,
 and the fragmentation time, respectively. 
The dotted lines denote the evolution of the temperature and 
 contraction time for the model without the three-body reactions.
The contraction and cooling times are
 defined as $t_{\rm dyn}\equiv \rho/\dot{\rho}$ and  
 $t_{\rm cool}\equiv 3nkT/(2\Lambda _{\rm net})$, respectively.
The fragmentation time is defined to be the inverse of the 
 growth rate of the fastest-growing linear perturbation, 
  $t_{\rm frag}\equiv 2.07 /\sqrt{2\pi G\rho_c}$
 (eq. [38] of Nakamura, Hanawa, \& Nakano 1993).
It should be noted that a filament does not undergo fragmentation
 in one-dimensional calculations, therefore $t_{\rm frag}$
should be regarded as a measure of the free-fall time.

During the contraction, the temperature stays nearly constant 
 at $T \sim 300- 500$ K because of the H$_2$ line cooling.
The H$_2$ number fraction also stays nearly constant
 at $x_{\rm H_2}= 10^{-4}\sim 10^{-3}$
 until the density reaches $n_c \sim 10^8$ cm$^{-3}$.
After that, the three-body reactions for the H$_2$ formation
 become dominant.
Accordingly, the H$_2$ number fraction steeply rises around
$n_c \sim 10^{8} $ cm$^{-3}$ and almost all the hydrogen atoms 
 become into H$_2$ molecules by the stage
 at which the density reaches 
 $n_c \sim 10^{11} $ cm$^{-3}$.
In contrast, for the model without the three-body reactions,
 the H$_2$ number fraction stays nearly constant at 
 $x_{\rm H_2}= 10^{-4}\sim 10^{-3}$ and the contraction 
 stops at the stage at which the density reaches
 $n_c \sim 10^9$ cm$^{-3}$.

The heating rate by H$_2$ formation also becomes appreciable 
 when the three-body reactions become dominant. 
Therefore, the temperature rises slowly
 after the density reaches $n_c\sim 10^9$ cm$^{-3}$.
Note that the H$_2$ line cooling rate is 
 always the most efficient during the contraction.

During the contraction, the contraction time nearly
 coincides with the cooling time until the central density reaches
 $2 \times 10^{11}$ cm$^{-3}$.
This indicates that the cloud collapses in a cooling time.
After the central density reaches $10^4$ cm$^{-3}$, 
 the contraction proceeds quasi-statically 
 because the contraction time becomes
 longer than the fragmentation time.
In other words, if there are fluctuations, 
 the cloud is expected to become unstable to fragmentation
 after the density reaches $10^4$ cm$^{-3}$.
Such evolution is explained as follows.
A density of $10^4$ cm$^{-3}$ 
 is almost comparable to a critical density of H$_2$, 
  $n_{\rm cr} \sim 10^{3-4}$ cm$^{-3}$, 
 beyond which LTE populations are achieved
 for the rotational levels of hydrogen molecules.
If the density is less than $n_{\rm cr}$, then 
 the H$_2$ line cooling rate is nearly proportional to the square
 of the density, $\Lambda \propto n^2$, and 
 the cooling time is thus inversely proportional to the density, 
 $t_{\rm cool} \propto n^{-1}$.
On the other hand, if the density is greater
 than $n_{\rm cr}$, then the H$_2$ line cooling rate is nearly
 proportional to the density, $\Lambda \propto n$ and 
 accordingly the cooling time is independent of the density.
Therefore, after the density reaches $n_{\rm cr}$, 
 the cooling time becomes longer than the fragmentation time 
 ($t_{\rm frag} \propto n^{-1/2}$)
 when the temperature is nearly constant.
(In fact, the cooling time is almost constant
 during $10^4$ cm$^{-3}\lesssim n_0\lesssim 10^7$ cm$^{-3}$.)
After the central density reaches $10^8$ cm$^{-3}$, 
 the H$_2$ abundance steeply increases 
 due to the three-body reactions and
 the radial contraction thus  accelerates again owing to 
 the enhanced H$_2$ line cooling.
(In contrast, for the model without three-body reactions, 
 the radial contraction stops when the central density
 reaches $\sim 10^9$ cm$^{-3}$ because of less effective H$_2$ 
 cooling.)
When the central density reaches $10^{11-12}$ cm$^{-3}$, 
 the cloud becomes optically thick to almost all the H$_2$ lines
 which significantly contribute to the total cooling rate. 
Consequently, the contraction essentially stops at that stage.

Such evolution is also related to the dynamical stability of self-gravitating
 clouds.
Assuming the polytropic relation of the equation of state
 ($P\propto \rho ^\gamma$), a hydrostatic cylindrical cloud is stable
 to radial contraction when $\gamma > \gamma _{\rm cr}=1$.
In the present model, the temperature is nearly constant during the
 contraction. 
In other words, the cloud is in a marginally stable state.
Thus, once the H$_2$ line cooling becomes less effective, the radial 
 contraction easily stops.
This is essentially different from that of a spherical cloud.
A spherical polytropic gas cloud can continue to collapse 
as long as $\gamma < \gamma _{\rm cr}=4/3$ even if the 
cooling rate becomes less effective (see \S 4.1 and Omukai \& Nishi 1998).

\subsection{A Dense Filament:
 A Dynamically Collapsing Filament (Model C6a)}

As a typical example of dense filaments,
 we show the evolution of model C6a 
 which has the initial parameters of 
 $n_0=10^6$ cm$^{-3}$, $T_0=400$ K, and $f=4$.
The electron and H$_2$ number fractions are initially set to 
$5 \times 10^{-5}$ and $10^{-4}$, respectively.
Figures 2a and 2b are the same as 
 those of Figures 1a and 1b,
 respectively, but for model C6a.

The evolution is qualitatively similar to that of model A1a, 
 although the temperatures are about two times higher
 during the contraction.
Since the H$_2$ formation rates by the three-body reactions
 are inversely proportional to the temperature, 
 the three-body reactions become more effective at the later stages
 than for model A1a ($n\gtrsim 10^9$ cm$^{-3}$).
The contraction time does not become longer than
 the fragmentation time until the cloud becomes optically thick 
 to the H$_2$ lines ($n_c\sim 4\times10^{12}$ cm$^{-3}$).
In other words, the contraction proceeds dynamically.
This is due to the effective cooling by the three-body reactions.
Therefore, the fragmentation is not expected to take place 
 until the cloud becomes optically thick to the H$_2$ lines 
 ($n_c\sim 10^{12-13}$ cm$^{-3}$).
Note that, in the model without the three-body reactions, 
 the contraction time becomes longer than the fragmentation time 
 after the central density reaches $\sim 10^{10}$ cm$^{-3}$, 
 and the contraction stops when the central density reaches
 $\sim 10^{11}$ cm$^{-3}$,
 i.e., before the cloud becomes optically thick to the H$_2$ lines.

\subsection{Summary of One-Dimensional Simulations}

We pursued the evolution for the other model parameters
tabulated in Table \ref{tab:model}.
We found that the evolution of all the models
 falls into either of the two types described above, 
 depending mainly on the initial density.
When the initial density is lower than
 $n_0 \lesssim 10^5$ cm$^{-3}$, 
 the contraction time becomes longer than the 
 fragmentation time before the three-body reactions 
 become effective ($n\sim 10^{8-9}$ cm$^{-3}$).
However, the contraction time does not exceed 
 the fragmentation time until the density becomes greater than
 $n_{\rm cr}$, beyond which LTE populations are achieved
 for the rotational levels of H$_2$ molecules.
On the other hand, when the initial density is greater than
 $n_0 \gtrsim 10^5$ cm$^{-3}$ and $f\gtrsim 3$,
 the contraction proceeds dynamically
 until the cloud becomes optically thick to 
 the H$_2$ lines ($n_c\sim 10^{12}$ cm$^{-3}$).
For all the models, the radial contraction essentially stops 
 at the stage at which the cloud becomes optically thick to 
 the H$_2$ lines which significantly contribute 
 to the total cooling rate.

From the numerical results of the one-dimensional simulations,
 the filament is likely to fragment into pieces
 during the stages at which 
 the central density is greater than $\sim 10^4$ cm$^{-3}$ 
 ($\gtrsim n_{\rm cr}$) and is less than $\sim 10^{12}$ cm$^{-3}$ 
 because the radial contraction is appreciably decelerated. 
When the filament has an initial density
 lower than $10^5$ cm$^{-3}$, 
 the fragmentation is expected to take place
 by the stages 
 at which the three-body reactions become effective.
On the other hand,
 when the filament has an initial density
 greater than $10^5$ cm$^{-3}$, 
 the fragmentation is expected to take place 
 at the stages at which the cloud becomes optically thick to 
 the H$_2$ lines ($n_c\sim 10^{12}$ cm$^{-3}$).

In the next section, we pursue the fragmentation processes
 with two-dimensional simulations and evaluate the masses
 of the fragments.

\section{Fragmentation of Filamentary Clouds}
\label{sec:result2}

In this section, to estimate the masses of the fragments,
 we explore the evolution of collapsing filamentary clouds 
 with density fluctuations
 by means of two-dimensional simulations. 

As shown below, for several models (particularly for models
with a large $f$), radial contraction proceeds to a great degree
 before the density fluctuations grow nonlinearly.
The spatial resolution thus becomes poor before 
 the cloud fragments into smaller clumps.
To resolve the fragmentation, we refine grids according to the
following procedure:  
(1) First, in the linear stage of density fluctuations, 
the cloud evolution for a model is pursued 
parallel by a one-dimensional calculation without 
density fluctuations and by a two-dimensional calculation with 
density fluctuations. 
The grid spacings used in a one-dimensional calculation
are taken to be ten times finer than those of a
two-dimensional calculation. 
The one-dimensional calculation monitors the radial 
density profiles fairly accurately, while the two-dimensional calculation
traces the growth of density fluctuations in the linear stage.
(2) When the mean density in the $z$-axis reaches a reference 
density $\rho _{\rm ref}$, which ensures the linear stage (typically
$\rho _{\rm ref} = 10^{3-4} \rho _0$),
the density fluctuations in the two-dimensional calculation
are Fourier-transformed to give 
the power spectrum of the fluctuations.
(3) Then, the grids in the two-dimensional calculation are refined
in the region of $0\le r\le 0.1 R_{\rm max}$ and $0\le z\le 0.1 Z_{\rm max}$
and the radial density profiles by the one-dimensional calculation are
mapped upon the refined grids. 
(4) Finally, the Fourier-transformed density fluctuations 
obtained above are added on the refined two-dimensional grids.
Then, the evolution of that model is pursued on 
the new computational domain.
The number of grids is $1024^2$ for usual cases and $2048^2$
for high-resolution cases.

\subsection{A Low-Density Filament (Model A4a)}

In this subsection, we show the evolution of model A4a 
as a typical example of less dense filaments.
Figure 3 shows
 the cross-sections of the cloud at four different stages.
This model has the initial parameters of 
 $n_0=10^4$ cm$^{-3}$, $T_0=400$ K, and $f=1.5$.
At the initial state, scale-invariant density fluctuations
($\nu=-1$) with an amplitude of $\delta \rho/\rho =0.1$
 were added.
The grid number was set to $2048^2$.

At the early stages, the density fluctuations do not
 grow appreciably in time because the contraction time
 is shorter than the fragmentation time
 [panel (b) of Fig. 3].
When the mean density in the $z$-axis exceeds
 $\sim 7 \times 10^6$ cm$^{-3}$, the contraction time
 becomes longer than the fragmentation time and 
 the density fluctuations begin to grow nonlinearly.
As a result, the filamentary cloud fragments into denser clumps 
 by the stages at which the mean density in the $z$-axis
 reaches $4 \times 10^7$ cm$^{-3}$
 [panels (c) and (d) of Fig. 3].
The masses of the clumps are estimated to be
 $\sim 250 M_\odot$.
The mean separation of the clumps is nearly equal to 
 0.24 pc which is comparable to the wavelength of the 
 fastest-growing perturbation at the stage at which 
 the mean density in the $z$-axis reaches $10^7$ cm$^{-3}$.

To see the structures of the clumps more quantitatively, 
 we show the density and temperature distributions
 in the $z$-axis in Figure 4.
At the early stages,
 the temperature stays nearly constant at $T\sim 500 $K.
After the mean density in the $z$-axis reaches $10^7$ cm$^{-3}$,
 the density fluctuations grow nonlinearly, and 
 dense prolate clumps form.
As the collapse proceeds, the central region of the clump
 becomes spherical.
In the clumps, the contraction is accelerated, and 
 the temperature rises slowly.
This acceleration is related to the dynamical stability of 
 self-gravitating clouds. 
A cylindrical polytropic ($P\propto \rho ^\gamma$) cloud is
 stable to radial contraction
 when $\gamma \gtrsim \gamma _{\rm cr}=1$.
Therefore, the primordial filament collapses quasi-statically 
 because the effective $\gamma$ is slightly greater
 than $\gamma _{\rm cr}$ (see Fig. 1).
On the other hand, for a spherical cloud, the critical value
 of $\gamma$ is equal to $\gamma _{\rm cr}=4/3$.
Accordingly, once the fragmentation takes place, 
 the clumps become unstable to dynamical contraction, 
 resulting in a temperature rise.
Such evolution is similar to that of 
 the spherical collapse of primordial clouds
 (e.g., Omukai et al. 1998; Omukai \& Nishi 1998).

The evolutions of other models with $n_0\lesssim 10^{5-6}$ cm$^{-3}$ or
 $f\lesssim 2$ are qualitatively similar to that of this model
 (e.g., models A1a $-$ A6a, B1a $-$ B6a, C1a $-$ C3a).
The fragmentation can take place before the stages at which the three-body reactions 
 become efficient and the radial contraction is reaccelerated.

\subsection{A Dense Filament (Model C6a)}

In this subsection, we show the evolution of model C6a as a 
 typical example of dense filaments.
Figure 5 shows the density and temperature
 distributions in the $z$-axis at three different stages.
This model has the initial parameters of 
 $n_0=10^6$ cm$^{-3}$, $T_0=400$ K, and $f=4$.
At the initial state, scale-invariant density fluctuations
($\nu=-1$) with an amplitude of $\delta \rho/\rho =0.1$
 were added.
As expected from the numerical results of
 the one-dimensional simulations, the radial contraction
 proceeds dynamically until the mean density in the $z$-axis
 reaches $4\times 10^{12}$ cm$^{-3}$.
When the mean density in the $z$-axis exceeds
 $4\times 10^{12}$ cm$^{-3}$, 
 the cloud becomes optically thick to the H$_2$ lines, and 
 the density fluctuations then begin to grow nonlinearly.
In this way, the cloud fragments into clumps.
The fragment mass is reduced down to $1-2M_\odot$ 
 owing to the high density of the filament.

The evolutions of other models with $n_0\gtrsim 10^{5-6}$ cm$^{-3}$ and
 $f\gtrsim 2$ are similar to that of this model
 (e.g., models C4a $-$ C6a).
In those models, the fragment masses take their minimum values of $1-2M_\odot$
 because the radial contraction proceeds dynamically until the stages at which 
 the H$_2$ lines become opaque.

\subsection{Typical Masses of the Fragments}

Figure 6 shows the distribution
 of the averaged fragment mass derived from the two-dimensional 
 simulations for the models with $\nu =-1$ and $\delta \rho/\rho =0.1$.
The abscissa and ordinate denote the initial central density 
 and the parameter $f$, respectively.
Since for all the models calculated
 in Figure 6, the initial temperatures 
 are taken to be constant at $T_0=400$ K, 
 the ordinate specifies the initial line mass.
The solid lines denote the contours of the averaged fragment mass.
As discussed in the Appendix, the primordial filaments are
 expected to form by cosmological pancake collapse
 and fragmentation.
For comparison, the line masses of such filaments
 with $l=l_{\rm eq}$, $l=1.5l_{\rm eq}$, and $l=2l_{\rm eq}$
 are shown by dashed lines in Figure  6
 (see eq.[\ref{eq:line mass}]). 
Here, $l_{\rm eq}$ is the line mass of the filament in 
 hydrostatic equilibrium and  
 is determined by the gas temperatures at which
 the cooling time balances with 
 the fragmentation time of the pancaking disk, where 
 the fragmentation time is defined as the inverse of the 
 growth rate of the fastest-growing linear perturbation 
 (Larson 1985).

The averaged fragment mass depends on the initial parameters.
For larger $f$ and/or higher initial density, the fragment mass is lower.
As expected from the one-dimensional simulations, 
 the fragmentation takes place during the stages at which
 $10^4$ cm$^{-3}$ $\la n_c \la$ $ 10^{12}$ cm$^{-3}$ 
 because the radial contraction proceeds quasi-statically.
Then, the maximum and minimum masses are estimated
 as $10^3$ M$_\odot$ and $1\sim 2$ M$_\odot$, respectively.
It is worth noting that these two masses are related to 
 the microphysics of H$_2$ molecules.
The former corresponds to the Jeans mass at the stage at which 
 the density reaches the critical density of H$_2$ molecules, 
 while the latter corresponds to the Jeans mass at the stage at which 
 the cloud becomes opaque to the H$_2$ lines.

There is a steep boundary
 in the fragment mass at $n_0 \sim 10^{5-6}$ cm$^{-3}$
 and $f\gtrsim 3$.
For the models with $n_0 \gtrsim 10^5$ cm$^{-3}$, 
 the fragment masses take their minimums at $1\sim 2$ M$_\odot$. 
On the other hand, for the models with $n_0 \lesssim 10^5$ cm$^{-3}$,
 they are greater than $\sim 10^2$ M$_\odot$.
This sensitivity in the fragment mass comes from 
 the rapid increase in H$_2$ abundance due to
 the three-body reactions.
As shown in Figure 1, 
 when the three-body reactions become effective
 ($n\gtrsim 10^8$ cm$^{-3}$), the radial contraction accelerates
 again because of the enhanced H$_2$ line cooling.
For models with a smaller initial density
 ($n_0\lesssim 10^5$ cm$^{-3}$),
 linear density fluctuations can 
 grow nonlinearly before the three-body reactions
 become dominant at $n\lesssim 10^{8-9}$ cm$^{-3}$.
On the other hand, for models with a denser initial density
 ($n_0\gtrsim 10^{5-6}$ cm$^{-3}$),
 the contraction time does not exceed the fragmentation time
 until the cloud becomes optically thick to the H$_2$ lines,
 $n\sim 10^{12}$ cm$^{-3}$.

Although the fragment mass also depends on the initial temperature, 
the effect of the initial temperature is identical to
 that of parameter $f$. 
This is because changing $T_0$ and/or $f$ corresponds
 to the change of the line mass (see equation [\ref{eq:line mass}]).
During the contraction, the cloud temperature is
 determined by the balance between the heating and cooling rates.
In our model, the main heating source is the compressional heating 
 by gravitational contraction, while the main cooling source is the H$_2$
 line transitions.
Even if we take higher or lower initial temperature, the
 cloud temperature settles immediately to an equilibrium value at which 
 the heating rate is equal to the cooling rate.
Furthermore, the equilibrium temperature depends only weakly on density 
 ($T_{\rm eq}\sim 300-800$ K for $n_0=10-10^6$ cm$^{-3}$).
Therefore, the ordinate of Figure 6 can be replaced by $T_0$ if a
 constant $f$ is adopted.

The fragment mass depends on the amplitude and power index
 of the density fluctuations. In Figure 7,
 the results with
$\delta=0.01$ are shown. This figure shows that the dependence
on the fluctuation amplitude is quite weak and that the fragment mass
is not basically altered. Changing power index $\nu$ leads to
a change of amplitude for the most unstable mode. It is found
that the fragment mass is insensitive to the choice of $\nu$ in a range
of $-2 \leq \nu \leq 0$.

When the initial H$_2$ abundance is as high as $10^{-3}$, the fragment
 mass is reduced by a few tens \% because of the lower temperatures
  although the minimum values of the fragment mass do not change.

\section{Implications for the IMF of Population III Stars}
\label{sec:IMF}

As shown in the previous section, 
 the primordial filaments fragment into dense clumps whose 
 masses are in the range of
 $1M_\odot \la M \la 10^3 M_\odot$.
The masses of the clumps depend on the initial model
 parameters, particularly the initial density.
The initial densities of the filaments are
 related to the initial conditions of the parent clouds.
In the Appendix, based on a CDM cosmology,
 we considered the formation processes of the filaments and 
 estimated plausible initial conditions.
From eq. (\ref{eq: the initial density}) of the Appendix, 
 the initial densities of the filaments are estimated as
 $10^3$ cm$^{-3}$ $\lesssim n_0 \lesssim 10^7$
 cm$^{-3}$ for 1$\sigma$ density fluctuations 
 with masses of $M\simeq 10^{6-8} M_\odot$
 ($10^4$ cm$^{-3}$ $\lesssim n_0 \lesssim 10^8$
 cm$^{-3}$ for 3$\sigma$ density fluctuations).
Then, the clump masses are evaluated as 
 $1M_\odot\lesssim M \lesssim 500M_\odot$ 
 for 1$\sigma$ density fluctuations
 ($1M_\odot\lesssim M \lesssim 250M_\odot$
 for 3$\sigma$ density fluctuations).
Here, we assumed that the radius of the parent disk
 ranges from $0.01r_{\rm vir}$ to $0.1r_{\rm vir}$, where
 $r_{\rm vir}$ denotes the virial radius of the parent cloud
 (see eq. [\ref{eq:virial radius}]).
The minimum radius corresponds to that of a
 rotationally supported disk with a spin parameter of $\lambda =0.05$.
The maximum radius is taken from the numerical results
 by Bromm et al. (1999) who followed the collapse of 
 3$\sigma$ 'top-hat' density fluctuations
 with a mass of $\sim 10^6$M$_\odot$. 
Their numerical simulations indicate that 
 by the epoch of filament formation, 
 the radius of the disk shrinks to $\sim 0.1r_{\rm vir}$,
 i.e., before the disk contracts to form a  rotationally supported disk, 
 fragmentation takes place.

The dense clumps are expected to be the sites
 of Population III star formation.
Recently, Larson (2000) argues that these clumps
 are not likely to fragment into many lower-mass objects 
 because their masses are nearly comparable
 to the Jeans mass at the epoch of fragmentation.
Actually, numerical simulations (e.g., Bromm et al. 1999) have shown that
 the fragmentation of collapsing Jeans-mass clumps is likely
 to be limited to the formation of binary or small multiple systems.

In the primordial gas, 
 most of the parent clump mass is expected to accrete onto
 the subclumps that will evolve into Population III stars
 because metal-free (dust-free) gas is impervious
 to strong radiation pressure
 (Omukai 1999, private communication).
Thus, the masses of Population III stars are
 anticipated to become comparable to the masses of the clumps.

As mentioned in the previous section, the dependence of the 
 clump mass on the initial density exhibits a step
 around $n_0 \sim 10^5$ cm$^{-3}$.
Then, the IMF of Population III stars is likely to be 
 low-mass deficient and double-peaked
 at $m_{\rm p1}=1-2M_\odot$ and $m_{\rm p2}={\rm a \
 few}\times 10-10^2M_\odot$.
[The first peak is consistent with the estimates by
 Uehara et al. (1996) and Nakamura \& Umemura (1999).
The clumps around the second peak have similar masses to 
 those obtained by Abel et al. (1998) and Bromm et al. (1999).]
The masses of the clumps probably increase by merging with themselves.
The resultant mass spectrum could have 
 two power-law-like components with different peaks
 of $m_{\rm p1}\sim 1-2 M_\odot$ and $m_{\rm p2}\sim {\rm a \
 few}\times 10-10^2M_\odot$.
It should be noted that the relative height of the
 first peak probably descends with time 
 compared to that of the second peak,
 because the initial densities of the filaments 
 decrease with time.
In other words, at higher redshifts or
 for higher $\sigma$ density fluctuations,
 the contribution of the lower-mass component is more
 significant in the IMF.

\section{Metal Enrichment}

As discussed in \S \ref{sec:IMF},
 Population III stars are expected to be low-mass deficient 
 compared to present-day stars and 
 their IMF is likely to be bimodal with peaks of 
 $m_{\rm p1}\sim 1-2M_\odot$ and $m_{\rm p2} \sim {\rm a \ few}\times 10-10^2M_\odot$.
In Figure 8,
 we show a schematic IMF of Population III stars.

If each component of the IMF is approximated by a simple power law 
 with a sharp cut-off, then the IMF of 
 Population III stars is expressed as 
\begin{equation}
 \frac{dN}{d\log m}=\frac{d(N_{\rm low}+N_{\rm high})}
 {d\log m} \; ,
\label{eq:IMF}
\end{equation}
where
\begin{eqnarray}
 \frac{dN_{\rm low}}{d\log m}&=&\left\{\begin{array}{ll}
 Am^{-\alpha} & {\rm for} \ m\ge m_{\rm p1} \\
 0 & {\rm for} \  m< m_{\rm p1} \end{array}\right.
\label{eq:IMF1} \\
 \frac{dN_{\rm high}}{d\log m}&=&\left\{\begin{array}{ll}
Bm^{-\beta} & {\rm for}\ m\ge m_{\rm p2} \\
0 & {\rm for}\  m< m_{\rm p2} \end{array}\right.
\label{eq:IMF2}
\end{eqnarray}
where the value of $m_{\rm p2}$
 is somewhat arbitrary because it depends on the initial
 densities of the filaments.
The relative heights of two peaks are also related to the
 power spectrum of the cosmological density fluctuations.
If the power indexes of the IMF are greater than unity 
 ($\alpha$ and $\beta>1$), then 
 the numerical constants $A$ and $B$ are approximated as 
\begin{equation}
 A\sim  (\alpha -1)\kappa \epsilon M_{\rm total}
 m_{\rm p1}^{\alpha-1}\; ,
\end{equation}
and
\begin{equation}
 B\sim (\beta -1)(1-\kappa)\epsilon M_{\rm total}
 m_{\rm p2}^{\beta -1} \; ,
\end{equation}
 respectively, where
 $M_{\rm total}$ is the parent cloud mass, 
 $\epsilon$ is the star-formation efficiency in the cloud, 
 and $\kappa$ is the ratio of the mass contained 
 in the lower-mass component to the total stellar mass.

According to recent theoretical studies on the evolution of
 metal-free stars, the massive metal-free stars with masses of
 (1) $10M_\odot\lesssim M_{\rm star} \lesssim 35M_\odot$
 and (2) $150M_\odot\lesssim M_{\rm star}\lesssim 250M_\odot$
 can enrich the intergalactic medium through supernova
 explosions (e.g., Heger et al. 2000).
Roughly speaking, the former star ejects about 10\% of its total mass ($\sim 1M_\odot$)
 as heavy elements through a supernova, while the latter
 ejects about 50\% ($\sim 75M_\odot$).

Then, the metallicity produced by Population III stars 
 can be estimated from the IMF (eq.[\ref{eq:IMF}]) as functions of 
 $\epsilon$, $\kappa$, $\alpha$, and $\beta$.
When the power indexes are tentatively assumed to be equal to each other
 ($\alpha = \beta$), the metallicity is evaluated as 
\begin{eqnarray}
 Z &\sim& \left[3.9\times 10^{-3}Z_\odot
 \left(\frac{\kappa}{0.5}\right)\right. \nonumber \\
 &&+ \left.2.2\times 10^{-2}Z_\odot
 \left(\frac{1-\kappa}{0.5}\right)
  \right]\left(\frac{\epsilon}{10^{-2}}\right) \nonumber
 \\&&\hspace{3cm} {\rm for} \ \alpha=\beta=1.35,\\
Z  &\sim&  \left[3.3\times 10^{-4}Z_\odot
 \left(\frac{\kappa}{0.5}\right)\right. \nonumber \\
 &&+ \left.9.3\times 10^{-3}Z_\odot
 \left(\frac{1-\kappa}{0.5}\right)
  \right]\left(\frac{\epsilon}{10^{-2}}\right) \nonumber
 \\&&\hspace{3cm} {\rm for} \ \alpha=\beta=3 \; ,
\end{eqnarray}
 where we took the second peak of $m_{\rm p2}=50M_\odot$.
When the contribution from the high-mass component is 
 more significant, the metallicity is larger.
If the star formation efficiency is as high as the
 present-day value of $\epsilon \sim 10^{-2}$ and 
 the power indexes are around $\alpha=\beta \sim 1.35-3$, 
 then the metallicity is estimated as $10^{-3}-10^{-2}Z_\odot$, 
 which is consistent with the metallicity $Z\lesssim 10^{-2}Z_\odot$
 observed in Ly$\alpha$ forest clouds with $N_{\rm HI} > 3\times 10^{14}$
 cm$^{-3}$ by Cowie et al. (1995) and Cowie \& Songaila (1998).
Thus, the heavy elements by the first enrichment may be
 responsible for the metallicity in the intergalactic medium
 at high redshifts.
This might be testable with observations of the metallicity
 and abundance ratios of heavy elements in those objects.

Furthermore, the first enrichment by Population III stars might play a
 significant role in the early evolution of galaxies
 or abundances of the intergalactic medium observed by
 X-ray (e.g., Zepf \& Silk 1996; Larson 1998).

\section{Conclusions}

We have explored the collapse and fragmentation of filamentary
 primordial gas clouds numerically, including the nonequilibrium processes
 for hydrogen molecule formation.
The simulations have shown that, depending upon the initial density,
 the evolution of the filaments can be classified into two types.
If a filament has relatively low initial density
such as $n_c \lesssim 10^5$ cm$^{-3}$, the radial contraction 
is slow due to less effective H$_2$ cooling, and it appreciably 
 decelerates at density higher than a critical density,
 where LTE populations are achieved
 for the rotational levels of H$_2$ molecules and 
 the cooling timescale becomes  accordingly longer
 than the free-fall timescale. 
This filament tends to fragment into dense clumps 
 before the central density reaches $10^{8-9}$ cm$^{-3}$
where H$_2$ cooling by three-body reaction is effective,
and the clump mass is more massive than some tens $M_\odot$. 
In contrast, if a filament is initially as dense
 as $n_c \gtrsim 10^5$ cm$^{-3}$, more effective H$_2$ cooling 
 with the help of three-body reaction allows the filament to contract 
 up to $n\sim 10^{12}$ cm$^{-3}$,
 for which the filament becomes optically thick to H$_2$ lines
 and then the radial contraction almost stops.
At this final hydrostatic stage, 
 the clump mass is lowered down to $\approx 1 - 2 M_\odot$
because of the high density of the filament. 
The dependence of clump mass upon
 the initial density could be translated into the dependence of the local
 amplitude of random Gaussian density fields or the epoch of collapse of
 a parent cloud.
Hence, the distribution of the clump mass
 predicts that the initial mass function of  
 Population III stars is likely to be bimodal
 with peaks of $\approx 1 - 2 M_\odot$
 and $\approx 10^2 M_\odot$, where the relative heights 
 could be a function of the collapse epoch.
At higher redshifts or
 for higher $\sigma$ density fluctuations,
 the contribution of the lower-mass component is likely to
 be more significant in the IMF and 
 the relative height of the first peak
 probably decreases with time
 because the initial densities of the filaments 
 are likely to descend with time.

In our model, we do not take the effects of external
 radiation into account. 
As suggested by Abel et al. (2000), a single star may form 
 at the central high-density region of
 the first collapsed low-mass objects.  
The radiative feedback from the very first stars 
might be significant for subsequent star formation in the surrounding medium
 (e.g., Ferrara 1998; Ciardi, Ferrara, \& Abel 2000).

We modeled the Population III IMF as a superposition of two
 power-law distributions with different peaks and  
 estimated the metallicity produced by Population III 
 stars in the first collapsed objects at high redshifts.
If the star formation efficiency is of the same order as the
 present-day value $\epsilon \sim 10^{-2}$, 
 then the metallicity is estimated as $10^{-3}-10^{-2}Z_\odot$. 
This metallicity is consistent with that observed
 in the intergalactic medium at high redshifts.

If a significant amount of stars with masses between a few $M_\odot$ and 
 8$M_\odot$ were formed,
 they might have evolved into white dwarfs until the present epoch.
The old white dwarfs might presently reside in the galactic
 halo and may be related to dark component observed
 in microlensing experiments,
 i.e., massive compact halo objects (MACHOs).
If the star formation efficiency is of the same order as 
 the present-day value, then the ancient white dwarfs 
 are likely to contribute a few tens \% of the dark mass
 in the galactic halo, which seems to be consistent with some
 constraints discussed by several authors (e.g., Charlot \&
 Silk 1995; Alcock et al. 2000; Lasserre et al. 2000;
 M\'endez \& Minniti 2000; Hodgkin et al. 2000; Ibata et al. 2000).

\acknowledgments

We are grateful to T. Nakamoto, R. Nishi, K. Omukai, 
 H. Susa, K. Tomisaka, and H. Uehara for valuable discussions.
Numerical computations were carried out on VPP300/16R and VX/4R 
 at the Astronomical Data Analysis Center
 of the National Astronomical Observatory, Japan and 
 workstations at the Center for Computational Physics,
 University of Tsukuba.
This work was financially supported in part by the Grant-in-Aid 
 for Scientific Research on Priority Areas 
 of the Ministry of Education, Science, Sports and Culture
[10147205 and 11134203 (FN) and 11640225 (MU)].

\appendix

\section{Formation of Filamentary Primordial Clouds}

In this appendix, we consider the formation process
 of filamentary primordial clouds 
 to estimate the typical values as the initial conditions.

We premise a gravitational instability scenario
 for the formation of cosmic structure.
A cosmological density perturbation larger than the Jeans scale 
 at the recombination epoch forms a flat pancake-like disk.
This process has been extensively studied
 by many authors (e.g., Zel'dovich 1970;
 Sunyaev \& Zel'dovich 1972; Cen \& Ostriker 1992a, 1992b;
 Umemura 1993).
Although the pancake formation was originally studied
 by Zel'dovich (1970) in the context of the adiabatic
 fluctuations in baryon or hot-dark-matter-dominated universes, 
 recent numerical simulations have shown that 
 such pancake structures also emerge in CDM cosmology
 (e.g., Cen et al. 1994).
Thus, the pancakes are thought to be a ubiquitous feature
 in gravitational instability scenarios.
In the following, we reconsider the pancaking
 of a cosmological density perturbation
 and fragmentation of the pancake into filamentary clouds.

We first suppose a spherical top-hat overdense region
 in the Einstein-de Sitter Universe.
The overdense region collapses due to self-gravity
 until a shock develops. 
After thermalization by the shock,
 the region forms a virialized system.
The virial radius and temperature of the overdense region
 are then estimated as (e.g., Padmanabhan 1993) 
\begin{equation}
r_{\rm vir}=1.9 \times 10^2 {\rm pc}
 \left({1+z_{\rm vir} \over 30} \right)^{-1} 
\left({M_b \over 10^{6} M_\odot} \right)^{1/3}
 \left({\Omega _b \over 0.1 } \right)^{-1/3}
h_{0.5} ^{-2/3} \; ,
\label{eq:virial radius}
\end{equation}
and
\begin{equation}
T_{\rm vir}=1.2\times 10^4 {\rm K}
\left({1+z_{\rm vir} \over 30} \right) 
\left({M_b \over 10^{6} M_\odot} \right)^{2/3}
\left({\Omega _b \over 0.1 } \right)^{-2/3}
h_{0.5}^{2/3} \  \; ,
\end{equation}
where $h_{0.5}$ is the Hubble constant
 in units of 50 km s$^{-1}$ Mpc$^{-1}$, 
 $z_{\rm vir}$ is the redshift epoch of virialization,
 $M_b$ is the baryonic mass,
 and $\Omega _b $ is the baryonic density parameter.

In a realistic situation, however, the overdense region
 is more or less aspherical. 
The deviation from spherical symmetry
 grows with time and a pancake disk consequently forms.  
After the shock is thermalized, the pressure force nearly
 balances the gravitational force
 in the vertical direction in the disk.
If the radiative cooling is effective,
 the temperature can descend to a lower value
 than the virial temperature. 
Recent numerical simulations 
 (e.g., Haiman et al. 1996; Abel et al. 1998; Bromm et al. 1999)
 have shown that when the first collapsed objects with masses
 of $10^{5-8}M_\odot$ are virialized,  
 H$_2$ relative abundance rises from its initial value of $10^{-6}$
 to $x_{\rm H_2}\sim 10^{-3}-10^{-4}$ and H$_2$ molecules then
 cool the gas to a temperature of $T\sim 300- 500$ K.
Consequently a thin baryonic disk forms,
 where the baryon density overwhelms that of the extended 
 virialized dark halo (e.g. Umemura 1993). 
Then we have a relation such as
$2\pi G\rho_{\rm d}H_{\rm d}^2\simeq c_s^2$, 
 where $\rho_{\rm d}$, $H_{\rm d}$, and $c_s$ denote 
 the characteristic density, thickness of the disk,
 and sound speed, respectively.
The density of the cooled disk is then estimated as
\begin{eqnarray}
n &=& {\rho_{\rm d} \over \mu} = {G \over 2 \pi k T} 
{M_b^2 \over R_{\rm d} {}^4} \\
     &=&  5.1 \times 10^4 \ {\rm cm ^{-3}}
\left({\alpha \over 0.1} \right) ^{-4}
\left({1+z_{\rm vir} \over 30} \right)^{4} 
\left({T \over 500 \, {\rm K}}\right) ^{-1}
  \left({M_b \over 10^{6} M_\odot} \right)^{2/3} 
\left({\Omega _b \over 0.1 } \right)^{4/3}
h_{0.5} {}^{8/3}  \; ,
\label{eq: the initial density}
\end{eqnarray}
 where $\mu$ is the mean molecular weight,
 $R_{\rm d}$ is the disk radius, and
 $\alpha (\equiv R_{\rm d}/r_{\rm vir})$
 is the disk radius in units of $r_{\rm vir}$.
Furthermore, an overdense region acquires angular momentum 
 through tidal spin-up by their surrounding fluctuations.
Then the overdense region cannot collapse into a disk smaller 
 than the centrifugal barrier.
The radius of the centrifugal barrier is given by  
\begin{equation}
r_{\rm barr} \simeq 0.01 r _{\rm vir} \left({\Omega _b \over 0.1}\right)^{-1}
\left({\lambda \over 0.05}\right)^2 \;
\end{equation}
 where $\lambda$ is a dimensionless spin parameter
$\lambda \equiv J|E|^{1/2}G^{-1}M^{-5/2}$
 (Sasaki \& Umemura 1996), which is peaked
 around $\lambda =0.05$ (Heavens \& Peacock 1988). 
The symbols $J$ and $E$ denote
 the total angular momentum and energy, respectively.
The radius $r_{\rm barr}$ gives a lower bound for the size
 of the first collapsed pancake,
 which leads to a condition as $\alpha \gtrsim 0.01$.

According to the linear theory of the disk
 in hydrostatic equilibrium, the disk is most unstable 
 to the perturbation with the wavelength of 
 $\lambda _{\rm max}=4\pi H_{\rm d}$ (e.g., Larson 1985).
When the most unstable perturbation grows in the disk, the 
 disk fragments into filamentary clouds rather than 
 spherical clouds.
Then, the separation of the filaments is equal to $\lambda _{\rm max}$.

If all the mass within one wavelength $\lambda _{\rm max}$ 
 collapses into one filament of line mass $l$,
 then the mass of the filament is given as 
\begin{equation}
M_{\rm fil}\sim lR_{\rm d}=
 8.3\times 10^4 M_\odot 
 \left(\frac{\alpha}{0.1}\right)
 \left(\frac{1+z_{\rm vir}}{30}\right)^{-1} 
 \left(\frac{T}{500{\rm K}}\right)
 \left({M_b \over 10^6 M_\odot} \right)^{1/3}
 \left({\Omega _b \over 0.1 } \right)^{-1/3}
 h_{0.5}^{-2/3} \  \; .
\end{equation}
The number of filaments formed from the parent disk is then 
 estimated as
\begin{equation}
 N_{\rm fil}=\frac{M_b}{M_{\rm fil}}\simeq 12 
 \left(\frac{\alpha}{0.1}\right)^{-1}
 \left(\frac{1+z_{\rm vir}}{30}\right)
 \left(\frac{T}{500{\rm K}}\right)^{-1}
 \left({M_b \over 10^6 M_\odot} \right)^{2/3}
 \left({\Omega _b \over 0.1 } \right)^{1/3}
 h_{0.5}^{2/3} \; .
\end{equation}

\renewcommand{\thefigure}{\arabic{figure}}
\renewcommand{\thetable}{\arabic{table}}

\begin{deluxetable}{llll}
\tablecolumns{4}
\tablecaption{REACTION RATE COEFFICIENTS}
\tablehead{
&\colhead{Reactions} & \colhead{Rate Coefficients$^{\rm a}$} 
& \colhead{Reference}}
\startdata
  (H1) & $ {\rm H}+ e \rightarrow {\rm H^+}+2 e  $
& $ k_{H1}=5.85 \times 10^{-11}  T^{0.5} \exp (-157809.1/T)
[1+(T/10^5)^{0.7}] ^{-1}$ & C92 \\
  (H2) &$ {\rm H^+}+e \rightarrow {\rm H} + h \nu  $
& $ k_{H2} = 8.40 \times 10^{-11} T^{-1/2} (T/10^3)^{-0.2} 
[1 + (T/10^6)^{0.7}] ^{-1}$ & C92 \\
  (H3) & $ {\rm H}+{\rm H}\rightarrow {\rm H}+
{\rm H^+}+e $ 
& $ k_{H3}=1.7 \times 10^{-4} k_1 $ & PSS83 \\
  (H4) &${\rm H}+e \rightarrow {\rm H ^-}+h\nu $
&  $k_{H4}=1.4 \times10^{-18} T^{0.928} 
\exp\left(-T/16200\right)$  & GP98 \\
  (H5) &$ {\rm H}+{\rm H^-}\rightarrow{\rm H _2}+e$ 
& $ k_{H5} =\left\{\begin{array}{ll}
1.5 \times 10^{-9}  & T\le 300{\rm K}\\
4.0 \times 10^{-9}T^{-0.17} & T>300 {\rm K}  \end{array}\right.
$ & GP98 \\
  (H6) & $ {\rm H }+{\rm H^+}\rightarrow {\rm H_2 ^+} 
+ h\nu $ & $k _{H6} = \begin{array}{l}{\rm dex} [-19.38 -1.523\log T \\
\hspace{1cm}+1.118(\log T)^2-0.1269(\log T)^3]\end{array}$& GP98 \\
  (H7) & ${\rm H_2 ^+}+{\rm H}\rightarrow {\rm H_2} + 
{\rm H^+}$ & $ k_{H7}= 6.4 \times 10^{-10} $ & GP98 \\
  (H8) & ${\rm H_2}+{\rm H}\rightarrow 3 {\rm H}$ &
 $ k_{H8} \; \mbox{[see eqs. (B6)$-$(B8) in reference]} $ & SK87 \\
  (H9) & ${\rm H_2}+{\rm H^+}\rightarrow {\rm H_2 ^+} 
+ {\rm H} $ 
& $k_{H9} = \left\{\begin{array}{ll}
3.0 \times 10^{-10} \exp\left(-21050/T\right)
 & T \le 10^4 {\rm K} \\
1.5 \times 10^{-10} \exp\left(-14000/T\right) & T>10^4 {\rm K} 
\end{array} \right. $ & GP98 \\
  (H10) & ${\rm H_2}+e\rightarrow {\rm H} + {\rm H^-} $
& $k_{H10} =2.7\times 10^{-8} T^{-1.27} \exp(-43000/T) $ & GP98 \\
  (H11) & ${\rm H_2}+e\rightarrow 2 {\rm H}+e $
& $k_{H11} =4.38 \times 10^{-10} T^{0.35} \exp
(-102000/T) $ & SK87 \\ 
  (H12) & ${\rm H_2}+{\rm H_2}\rightarrow 2 {\rm H}+{\rm H_2} $ & $k_{H12} 
\; \mbox{[see eqs. (B6)$-$(B8) in reference]} $ & SK87 \\
  (H13) & ${\rm H^-}+e \rightarrow {\rm H}+2e$
& $ k_{H13}=4.0 \times 10^{-12} \exp (-43000/T) $ & SK87 \\
  (H14) & ${\rm H^-}+{\rm H}\rightarrow 2 {\rm H}+e $ 
& $k_{H14}=5.3 \times 10^{-20} T^{2.17} \exp (-8750/T) $ & SK87 \\
  (H15) & ${\rm H^-}+{\rm H^+}\rightarrow 2{\rm H} $ 
& $ k_{H15} =\begin{array}{l}5.7 \times 10^{-6} T^{-0.5}
+6.3\times 10^{-8} \\
\hspace{1cm}-9.2\times 10^{-11} T^{0.5}
+4.4 \times10^{-13}T\end{array}$ & GP98 \\
  (H16) & ${\rm H^-}+{\rm H^+} \rightarrow {\rm H_2 ^+}
+ e $ & $k_{H16} = \left\{\begin{array}{ll}
6.9 \times 10^{-9} T^{-0.35} & T \le 8000 {\rm K} \\
9.6 \times 10^{-7} T^{-0.9} & T>8000 {\rm K} 
\end{array} \right. $ & GP98 \\
  (H17) & ${\rm H_2 ^+}+e\rightarrow 2 {\rm H} $
& $ k_{H17} = 2.0 \times 10^{-7} T^{-0.5} $ & GP98 \\
  (H18) & $ {\rm H_2 ^+}+{\rm H^-} \rightarrow {\rm H}+{\rm H_2} $ 
& $ k_{H18}=5.0 \times 10^{-6} T^{-0.5}$ & SK87 \\
  (H19) & $ 3 {\rm H} \rightarrow {\rm H_2}+{\rm H} $ 
& $ k_{H19}=5.5 \times 10^{-29} T^{-1} $ & PSS83 \\
  (H20) & $2 {\rm H}+{\rm H_2} \rightarrow 2 {\rm H_2} $
& $ k_{H20}=k_{H19} / 8 $ & PSS83 \\
  (He1) & ${\rm He}+e \rightarrow {\rm He ^+}+2e$
&$ k_{He1}=2.38 \times 10^{-11} T ^{-0.5} \exp (-285335.4 /T)
[1+(T/10^5)^{0.5}]^{-1} $ & C92 \\
  (He2) & ${\rm He^+}+e \rightarrow {\rm He^{++}}+2 e $
& $ k_{He2}=5.68 \times 10^{-12} T^{0.5} \exp (-631515.0 /T) 
[1+(T/10^5)^{0.5}]^{-1} $ & C92 \\
  (He3) & $ {\rm He ^+}+e \rightarrow {\rm He}+h \nu $
& $  k_{He3}= \begin{array}{l} 3.294\times 10^{-11} 
\left[\sqrt{\frac{T}{15.54}}
 \left(1+\sqrt{\frac{T}{15.54}}\right)^{0.309} \right.\times \\
\left. \left(1+\sqrt{\frac{T}{3.676\times 10^7}}\right)^{1.691}\right]^{-1}
 \end{array} $ & VF96 \\ 
  (He4) & ${\rm He ^{++}}+e \rightarrow {\rm He ^+}+h \nu $
& $ k_{He4}=\begin{array}{l} 1.891\times 10^{-10} 
\left[\sqrt{\frac{T}{9.37}}
\left(1+\sqrt{\frac{T}{9.37}}\right)^{0.2476} \right.\times \\
\left. \left(1+\sqrt{\frac{T}{2.774\times 10^6}}\right)^{1.7524}\right]^{-1}
\end{array}$ & VF96   \\
  (He5) & ${\rm He}+ {\rm H^+} \rightarrow {\rm He^+}+{\rm H}$
& $k_{He5}=4.0 \times 10^{-37}T^{4.74}$ & GP98 \\
  (He6) & ${\rm He^+}+ {\rm H} \rightarrow {\rm He}+{\rm H^+}$
& $k_{He6}=3.7 \times 10^{-25}T^{2.06} 
\left[1+9.9\exp \left(-\frac{T}{2570}\right)\right]$ & GP98 \\
\enddata
\tablenotetext{a}{The units of rate coefficients are taken to be
cm$^3$ s$^{-1}$ for two-body reactions and cm$^6$ s$^{-1}$ for 
three-body reactions.}
\tablecomments{
C92: Cen (1992); PSS83: Palla et al. (1983); 
GP98: Galli \& Palla (1998); SK87: Shapiro \& Kang (1987); 
VF96: Verner \& Ferland (1996)}
\label{tab:1}
\end{deluxetable}

\begin{deluxetable}{lccccc}
\tablecolumns{5}
\tablewidth{.5\columnwidth}
\tablecaption{MODEL PARAMETERS}
\tablehead{
\colhead{Model} & \colhead{$f$} & \colhead{$n_c$ (cm$^{-3}$)}
 & \colhead{$T$ (K)}  & \colhead{$x_{\rm H_2}$}} 
\startdata

A1a  & 1.5 & $10$ & 400 & $10^{-4}$ \\
A1b  & 1.5 & $10$ & 300 & $10^{-3}$ \\
A2a  & 1.5 & $10^2$ & 400 & $10^{-4}$ \\
A2b  & 1.5 & $10^2$ & 300 & $10^{-3}$ \\
A3a  & 1.5 & $10^3$ & 400 & $10^{-4}$ \\
A3b  & 1.5 & $10^3$ & 300 & $10^{-3}$ \\
A4a  & 1.5 & $10^4$ & 400 & $10^{-4}$ \\
A4b  & 1.5 & $10^4$ & 300 & $10^{-3}$ \\
A5a  & 1.5 & $10^5$ & 400 & $10^{-4}$ \\
A5b  & 1.5 & $10^5$ & 300 & $10^{-3}$ \\
A6a  & 1.5 & $10^6$ & 400 & $10^{-4}$ \\
A6b  & 1.5 & $10^6$ & 300 & $10^{-3}$ \\
B1a  & 2 & $10$ & 400 & $10^{-4}$ \\
B1b  & 2 & $10$ & 300 & $10^{-3}$ \\
B2a  & 2 & $10^2$ & 400 & $10^{-4}$ \\
B2b  & 2 & $10^2$ & 300 & $10^{-3}$ \\
B3a  & 2 & $10^3$ & 400 & $10^{-4}$ \\
B3b  & 2 & $10^3$ & 300 & $10^{-3}$ \\
B4a  & 2 & $10^4$ & 400 & $10^{-4}$ \\
B4b  & 2 & $10^4$ & 300 & $10^{-3}$ \\
B5a  & 2 & $10^5$ & 400 & $10^{-4}$ \\
B5b  & 2 & $10^5$ & 300 & $10^{-3}$ \\
B6a  & 2 & $10^6$ & 400 & $10^{-4}$ \\
B6b  & 2 & $10^6$ & 300 & $10^{-3}$ \\
C1a  & 4 & $10$ & 400 & $10^{-4}$ \\
C1b  & 4 & $10$ & 300 & $10^{-3}$ \\
C2a  & 4 & $10^2$ & 400 & $10^{-4}$ \\
C2b  & 4 & $10^2$ & 300 & $10^{-3}$ \\
C3a  & 4 & $10^3$ & 400 & $10^{-4}$ \\
C3b  & 4 & $10^3$ & 300 & $10^{-3}$ \\
C4a  & 4 & $10^4$ & 400 & $10^{-4}$ \\
C4b  & 4 & $10^4$ & 300 & $10^{-3}$ \\
C5a  & 4 & $10^5$ & 400 & $10^{-4}$ \\
C5b  & 4 & $10^5$ & 300 & $10^{-3}$ \\
C6a  & 4 & $10^6$ & 400 & $10^{-4}$ \\
C6b  & 4 & $10^6$ & 300 & $10^{-3}$ \\
\enddata
\label{tab:model}
\end{deluxetable}

\clearpage
\begin{figure}
\plottwo{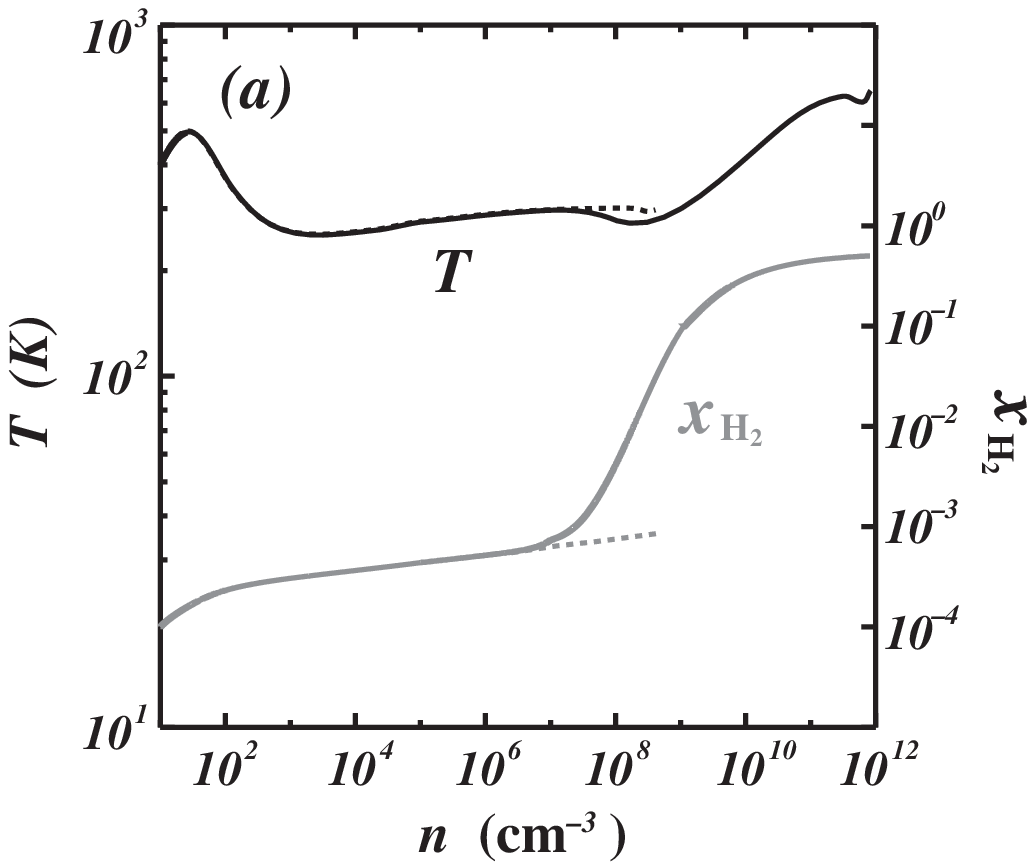}{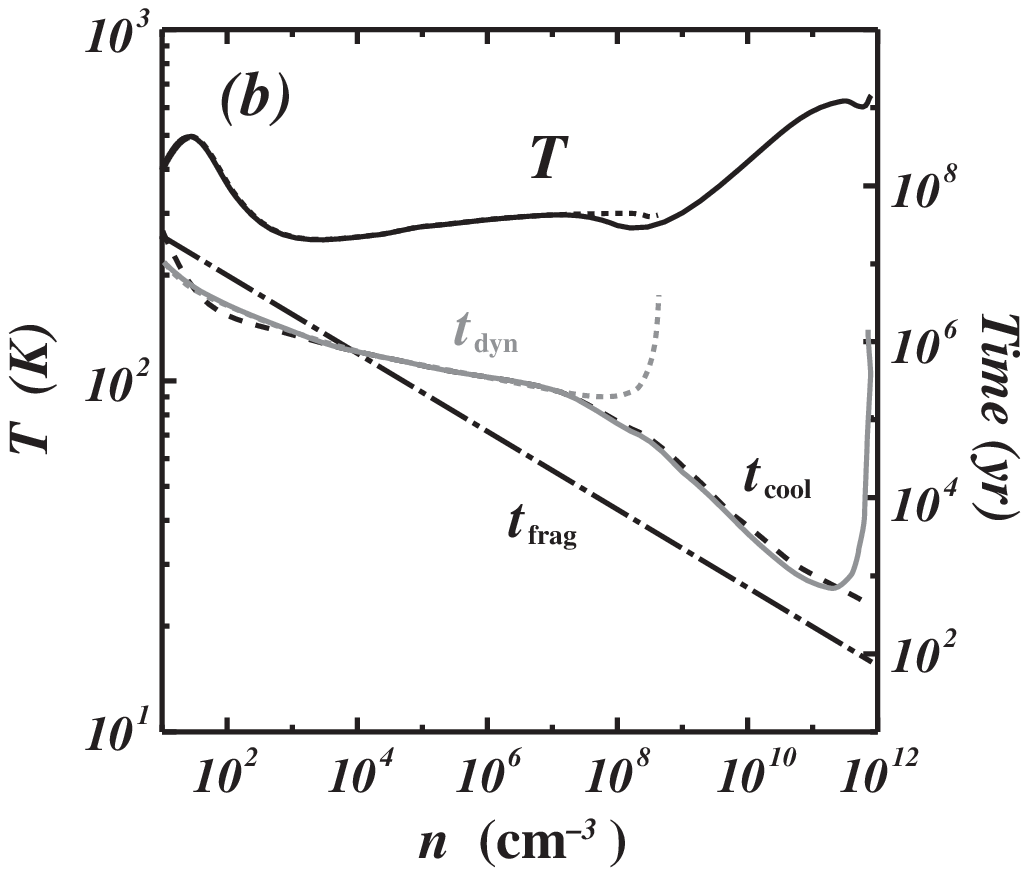}
\caption{(a) 
Evolution of the temperature ($T_c$) and the H$_2$
 relative abundance ($x_{\rm H_2, c}$) at the center 
 for model A1a with no density fluctuations.
The model has the initial parameters of 
 $(n_0, T_0, f)=(10 \, {\rm cm}^{-3},\ 400 \, {\rm K}, \ 1.5)$. 
For comparison, we also show the evolutional paths of the model 
 in which the three-body reactions are neglected
 ({\it dotted lines}).
The temperature stays nearly constant at $300-500$ K
 over the seventh order of magnitude in density.
When the central density reaches $n_c\sim 10^8$ cm$^{-3}$, 
 the three-body reactions of H$_2$ formation become dominant
 and almost all the hydrogen atoms become into H$_2$ molecules.
Thus, the H$_2$ abundance steeply rises around
 $n_c \sim 10^{8} $ cm$^{-3}$.
When the density reaches $n_c \sim 10^{12} $ cm$^{-3}$,
 the cloud becomes optically thick to the H$_2$ lines.
(b) Evolution of the contraction time ({\it solid line}),
 the cooling time ({\it dashed line}), and 
 the fragmentation time ({\it dashed-dotted line}).
For comparison, we also showed the evolution of the contraction time 
 for the model in which the three-body reactions are neglected
 ({\it dotted lines}).
The contraction becomes slower 
 after the central density reaches the critical density of H$_2$ 
 ($n_{\rm cr}\sim 10^{3}$ cm$^{-3}$).
When the density reaches $n_c \sim 10^{12} $ cm$^{-3}$, 
 the cloud becomes optically thick to the H$_2$ lines,
 and the contraction essentially stops at that stage.
For the model without the three-body reactions, 
 the contraction decelerates at the stage at which 
 $n_c \sim 10^{9} $ cm$^{-3}$
 because of less effective H$_2$ cooling.}
\label{fig:n-t}
\end{figure}

\begin{figure}
\epsscale{0.6}
\plottwo{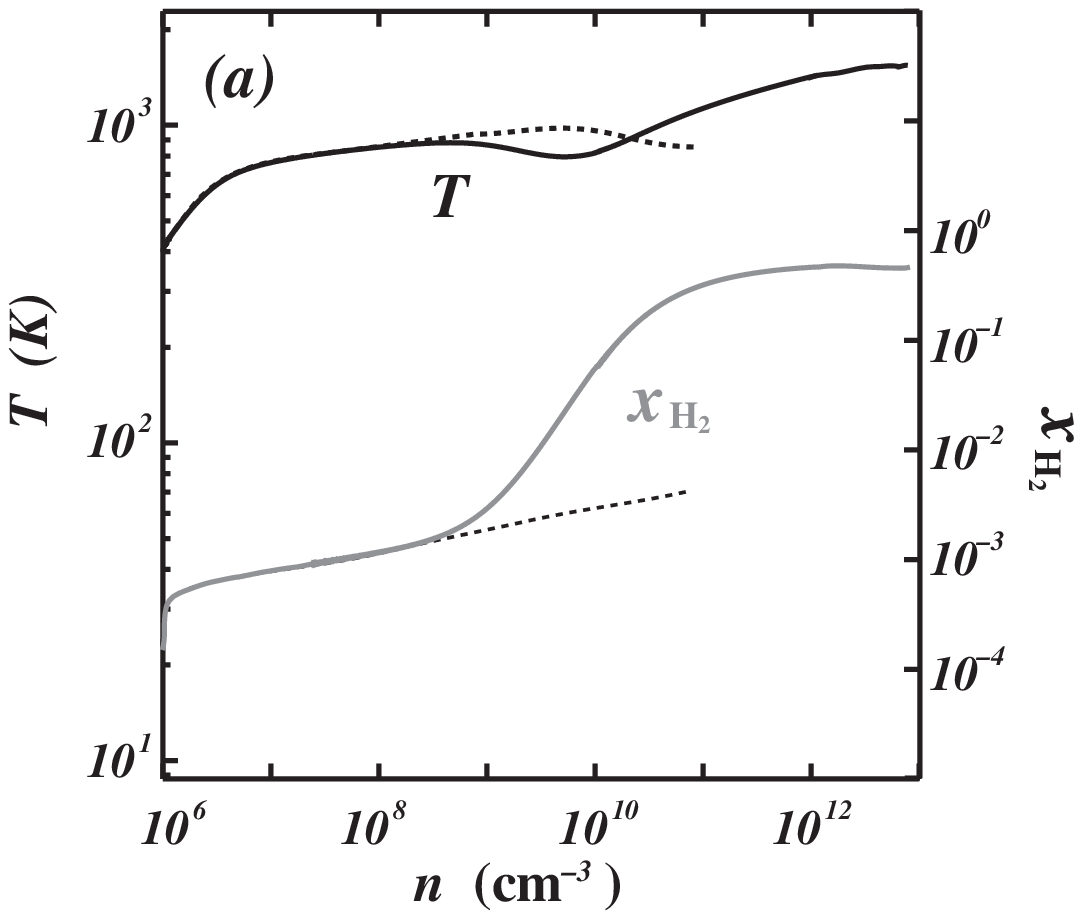}{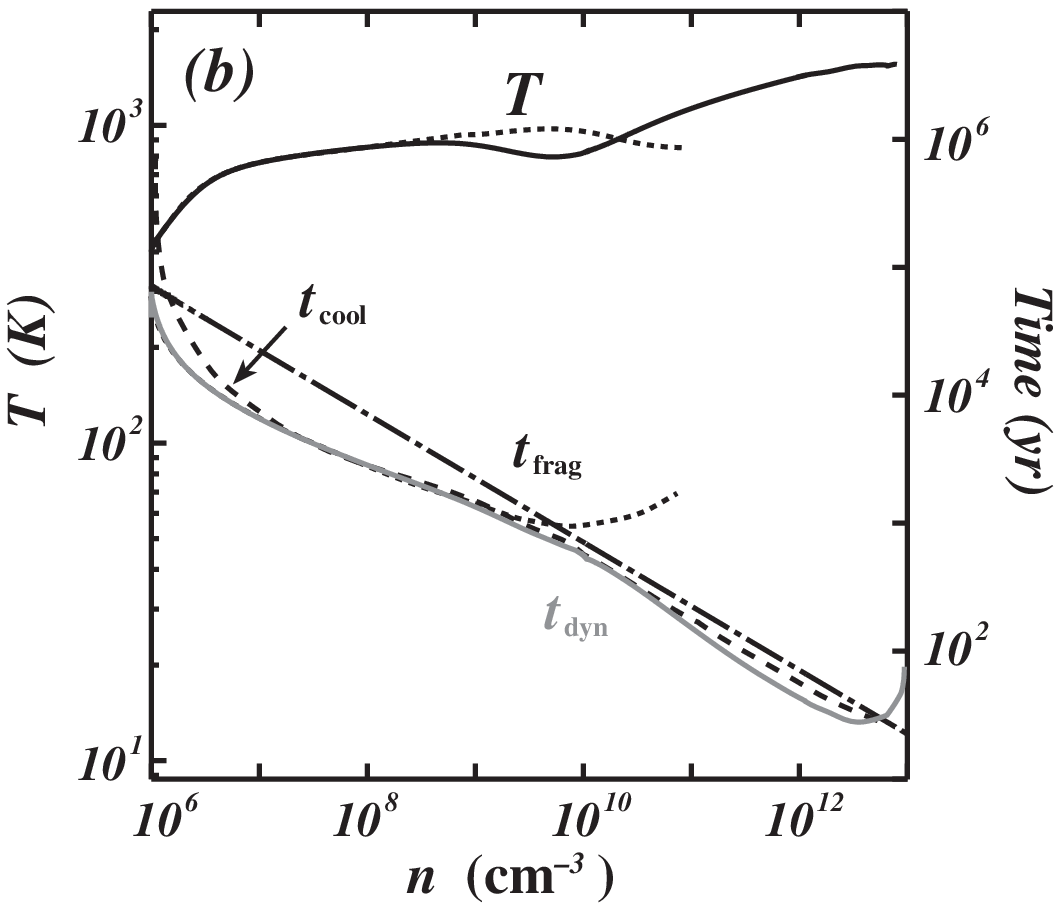}
\caption{
Same as Figure 1 but for model C6a.
The model has the initial parameters of 
 $(n_0, T_0, f)=(10^6 \, {\rm cm}^{-3},\ 400 \, {\rm K}, \ 4)$. 
The overall evolution is qualitatively similar to
 that of model A1a.
In this model, the contraction time does not exceed the fragmentation time until 
 the cloud becomes optically thick to the H$_2$ lines and 
 the contraction essentially stops.  
\label{fig:n-t2}
}
\end{figure}

\begin{figure}
\epsscale{0.6}
\plotone{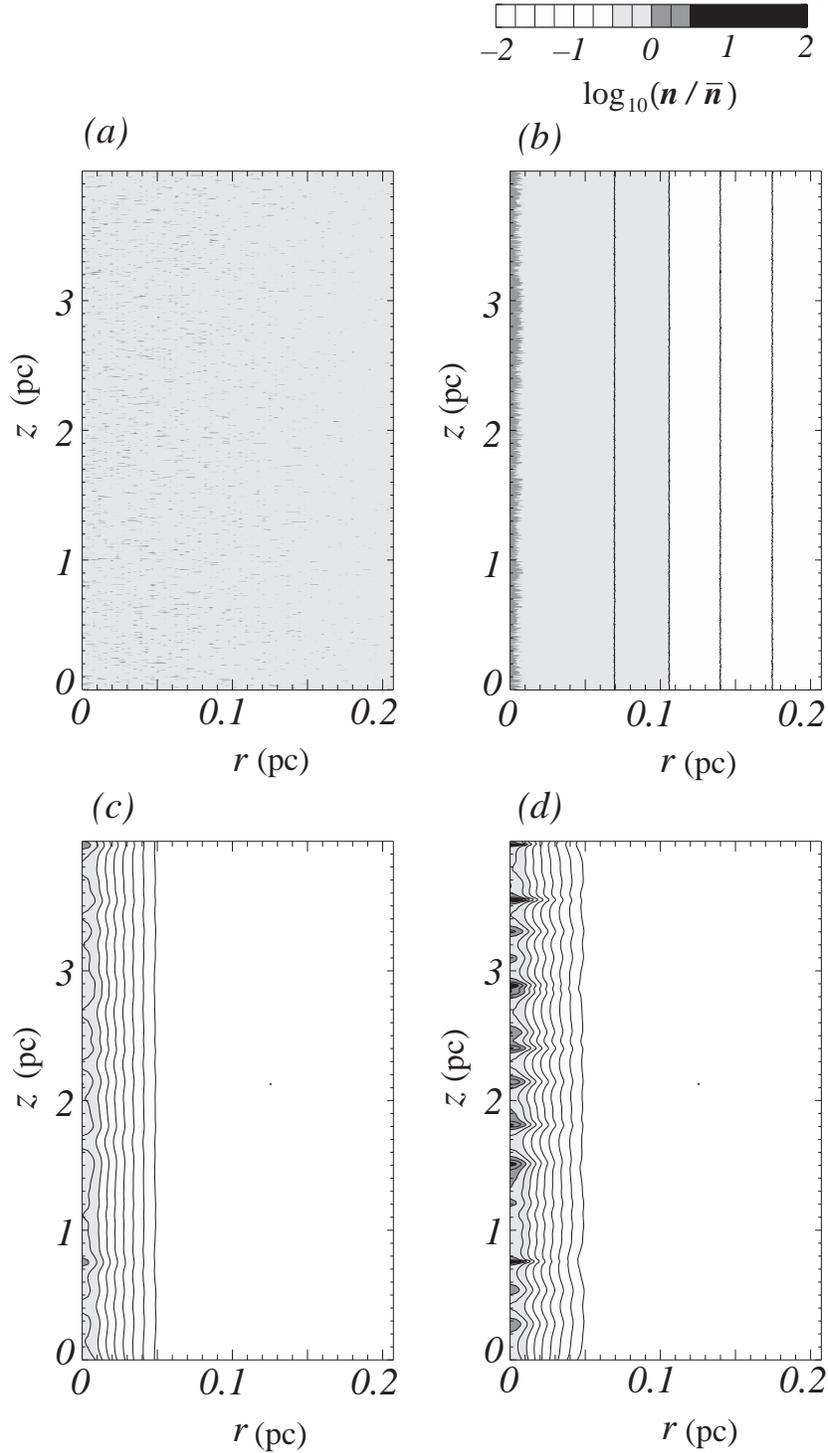}
\caption{Density and velocity distributions
 in the $r-z$ plane for model A4a at four different stages:
 (a) $t=0$ ($\bar{n_c} =1.0 \times 10^4$ cm$^{-3}$,
 initial state),
 (b) $t=5.9\times 10^5$ yr
 ($\bar{n_c} =9.9 \times 10^5$ cm$^{-3}$),
 (c) $t=8.1\times 10^5$ yr
 ($\bar{n_c} =5.1 \times 10^7$ cm$^{-3}$), 
 and (d) $t=8.2\times 10^5$ yr
 ($\bar{n_c} =7.5 \times 10^7$ cm$^{-3}$). 
At the initial state, scale-invariant density
 fluctuations with an amplitude of $\delta =0.1$ are added.
When the density reaches $10^7$ cm$^{-3}$, unstable density 
 fluctuations begin to grow nonlinearly because
 the contraction time becomes longer than the fragmentation time.
The mean mass and separation of the clumps are estimated
 as $\sim 250M_\odot$ and 0.24 pc at the final stage of the
 computation, respectively.
}
\label{fig:cross sections}
\end{figure}

\begin{figure}
\plottwo{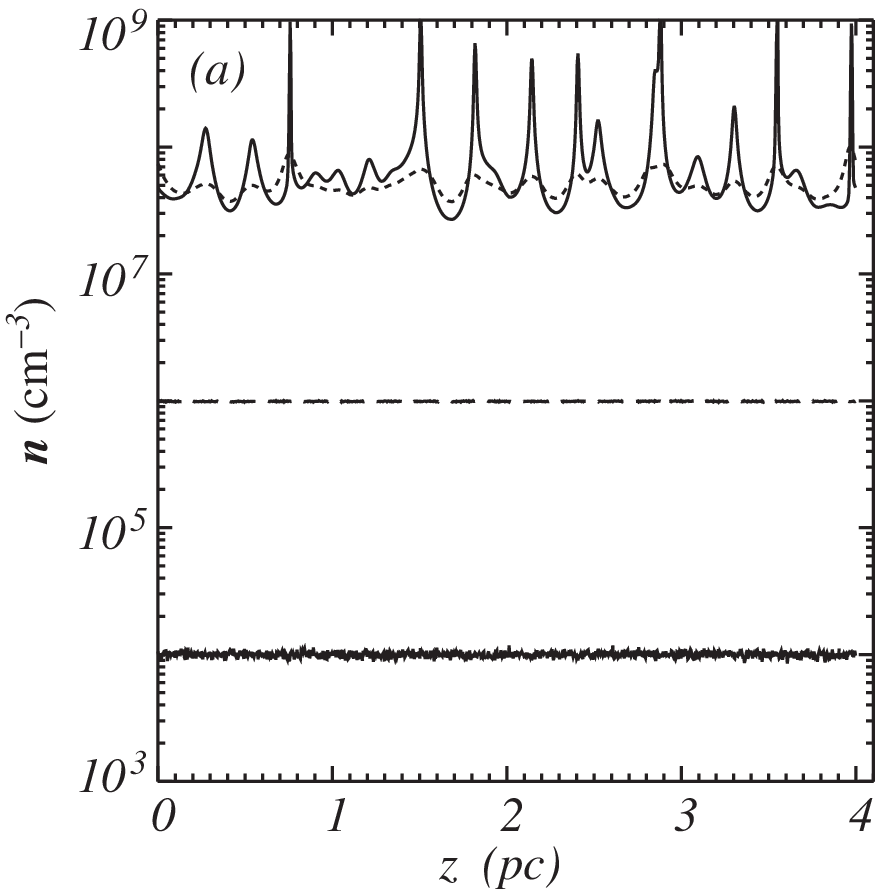}{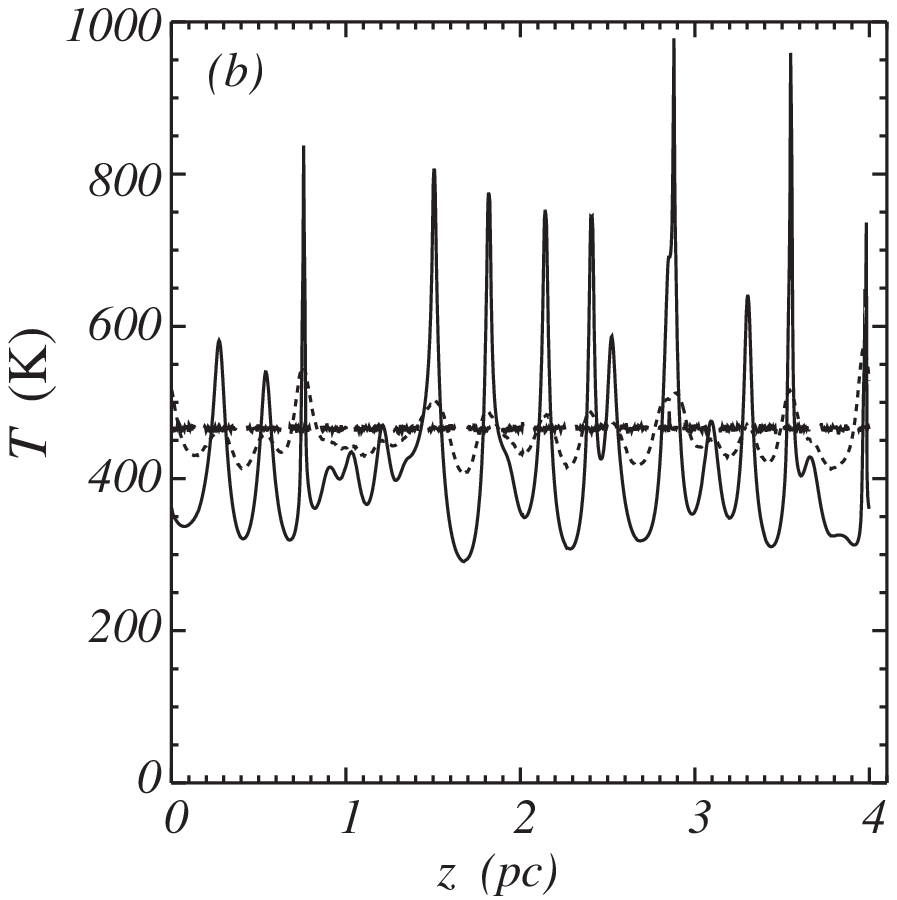}
\epsscale{0.6}
\caption{Evolution of (a) the density and (b) velocity
 distributions in the $z$-axis for model A4a
 with scale-invariant density fluctuations of $\delta =0.1$.
They are depicted at the stages at which 
 $t=0$ ({\it solid lines}),
 $t=5.9\times 10^5$ yr ({\it long dashed lines}),
 $t=8.1\times 10^5$ yr ({\it short dashed lines}), 
 and $t=8.2\times 10^5$ yr ({\it solid lines}).
When the density reaches $10^7$ cm$^{-3}$,
 the density fluctuations begin to grow nonlinearly, and 
 dense clumps form.
In the clumps, the contraction is accelerated,
 and the temperature accordingly rises.
}
\label{fig:profiles}
\end{figure}

\begin{figure}
\epsscale{0.6}
\plottwo{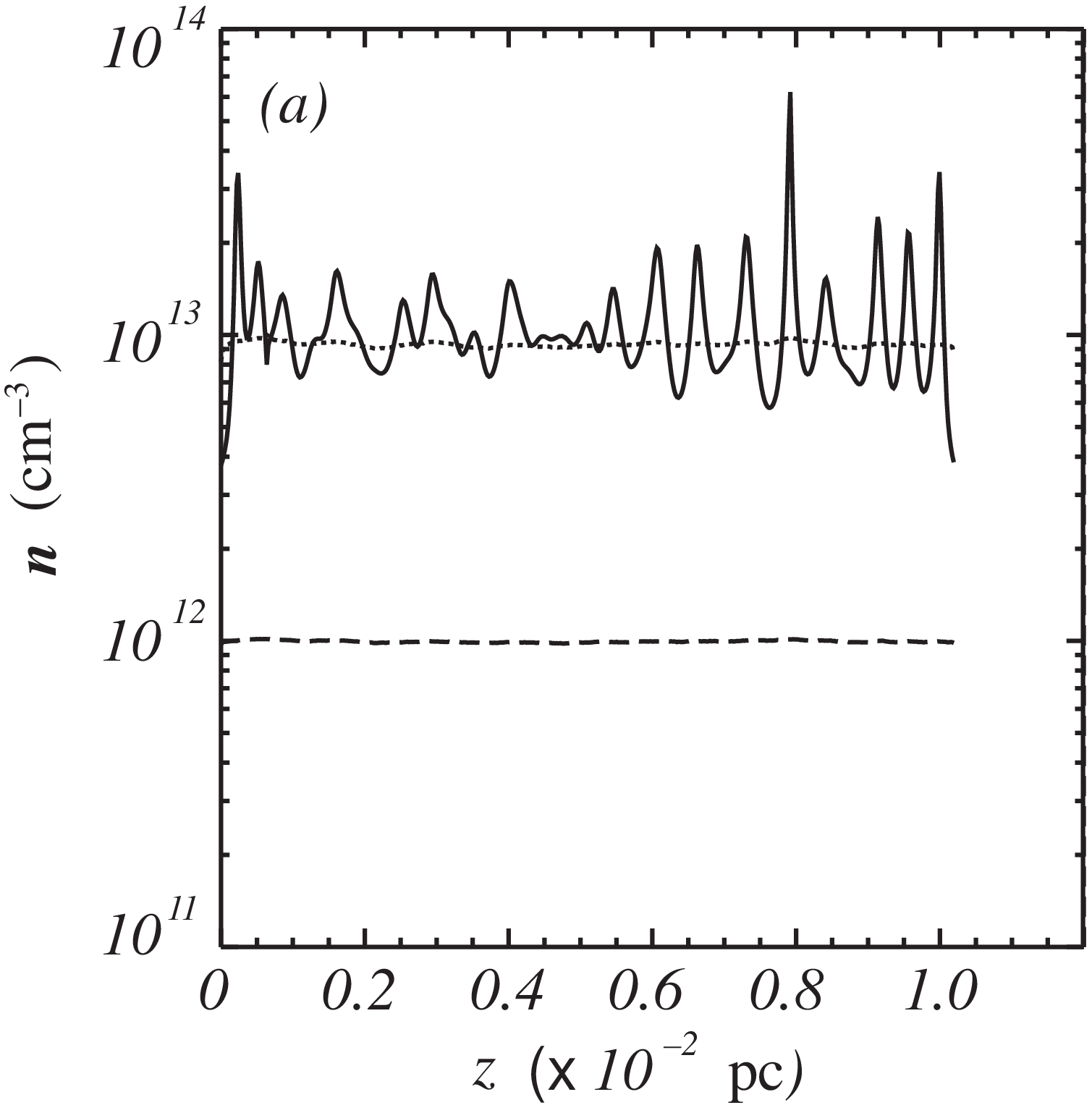}{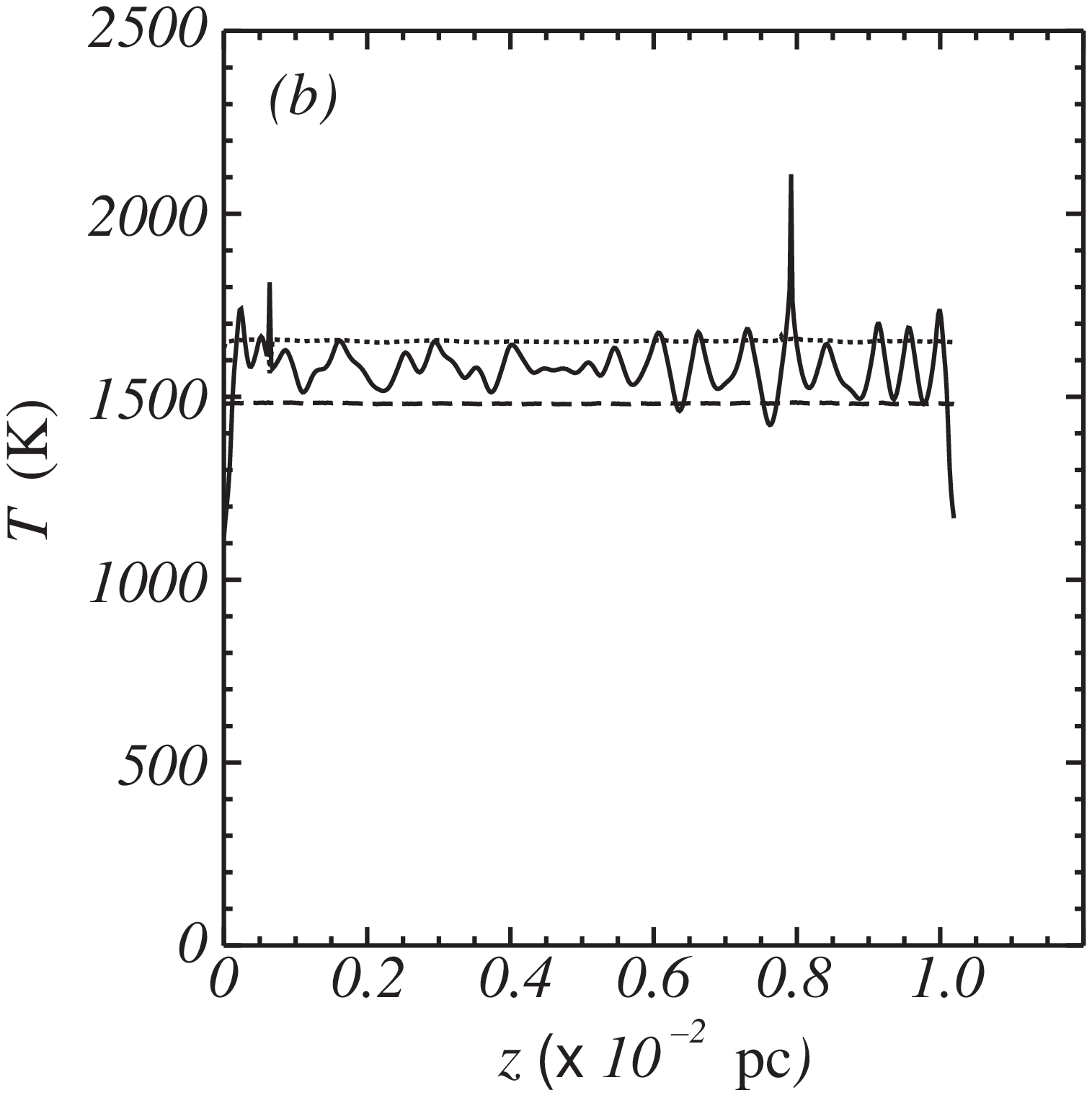}
\caption{Evolution of (a) the density and (b) velocity
 distributions in the $z$-axis for model C6a
 with scale-invariant density fluctuations of $\delta =0.1$.
They are depicted at the stages at which 
 $t=5.458\times 10^4$ yr ({\it long dashed lines}),
 $t=5.464\times 10^4$ yr ({\it dotted lines}), 
 and $t=5.47\times 10^4$ yr ({\it solid lines}).
When the density reaches $\sim 4\times10^{12}$ cm$^{-3}$,
 the radial contraction essentially stops; then, 
 the density fluctuations begin to grow nonlinearly, and 
 dense clumps form.
}
\label{fig:profiles2}
\end{figure}

\begin{figure}
\epsscale{0.6}
\plotone{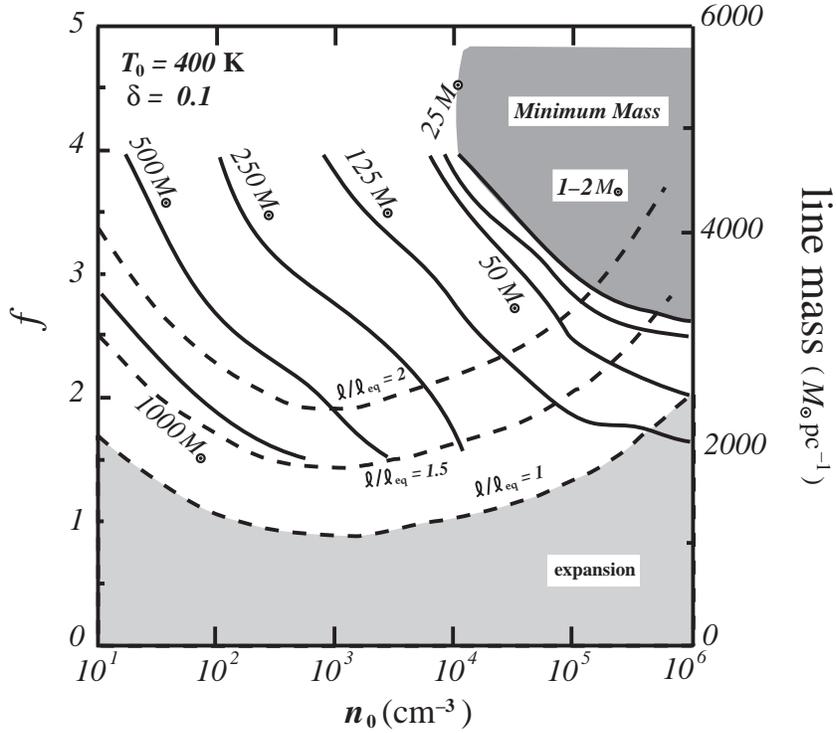}
\caption{
Dependence of fragment mass on the initial central density of 
 a filament $n_0$ and a parameter $f$ which  corresponds to the 
 initial line mass shown by the right ordinate.
Scale-invariant ($\nu =-1$) density fluctuations with an amplitude
 of $\delta=0.1$ are assumed.
Solid lines denote the contour lines of the averaged fragment mass.
Dashed lines show the line masses of the filaments formed 
 by cosmological collapse and fragmentation (see text and the Appendix).
The numbers attached to the dashed lines denote the line masses 
 normalized with the equilibrium value
 (e.g., $l/l_{\rm eq}=1$, 1.5, and 2).
When the line mass of the filament is less than 
$l/l_{\rm eq}=1$, the cloud expands because
 the pressure force is greater than the gravitational force
 ({\it light-gray region}).
The averaged fragment mass depends on the initial parameters.
For larger $f$ and/or higher initial density, the fragment mass is lower.
The maximum and minimum masses are estimated
 as $10^3$ M$_\odot$ and $1\sim 2$ M$_\odot$, respectively.
These two masses are related to the microphysics of H$_2$ molecules.
The former corresponds to the Jeans mass at the stage at which 
 the density reaches the critical density of H$_2$ molecules, 
 while the latter corresponds to the Jeans mass at the stage at which 
 the cloud becomes opaque to the H$_2$ lines.
There is a steep boundary
 in the fragment mass at $n_0 \sim 10^{5-6}$ cm$^{-3}$
 and $f\gtrsim 3$.
For the models with $n_0 \gtrsim 10^5$ cm$^{-3}$, 
 the fragment masses take their minimum at $1\sim 2$ M$_\odot$, 
 while for the models with $n_0 \lesssim 10^5$ cm$^{-3}$,
 they are greater than $\sim 10^2$ M$_\odot$.
This sensitivity in the fragment mass comes from 
 the rapid increase in the H$_2$ abundance due to
 the three-body reactions.
}
\label{fig:fragment mass}
\end{figure}

\begin{figure}
\epsscale{0.6}
\plotone{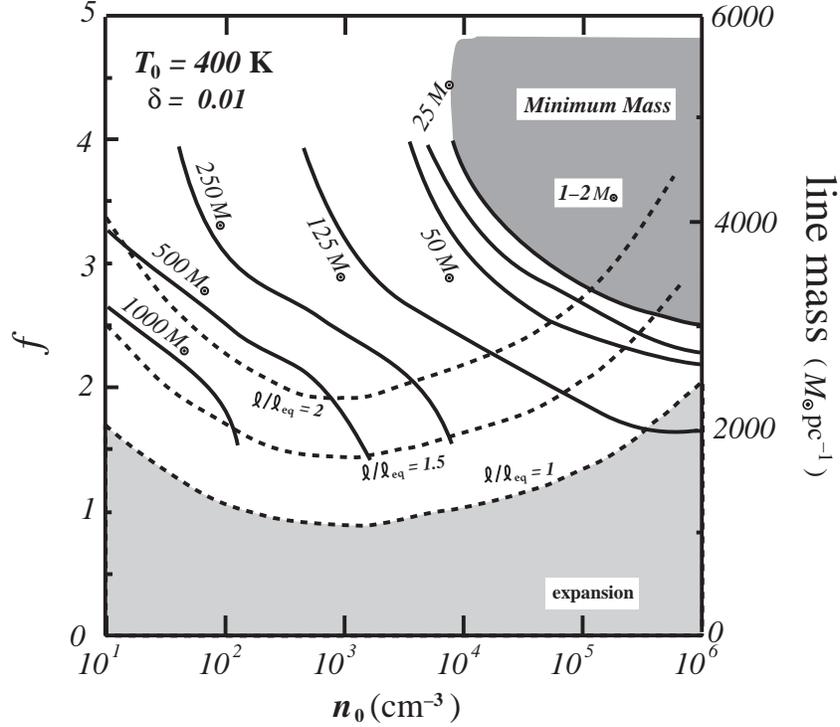}
\caption{
Same as Figure 6 but for $\delta =0.01$.
This figure shows that the dependence on the fluctuation amplitude
 is quite weak and the fragment mass is not basically altered. 
}
\label{fig:fragment mass2}
\end{figure}

\begin{figure}
\epsscale{0.6}
\plotone{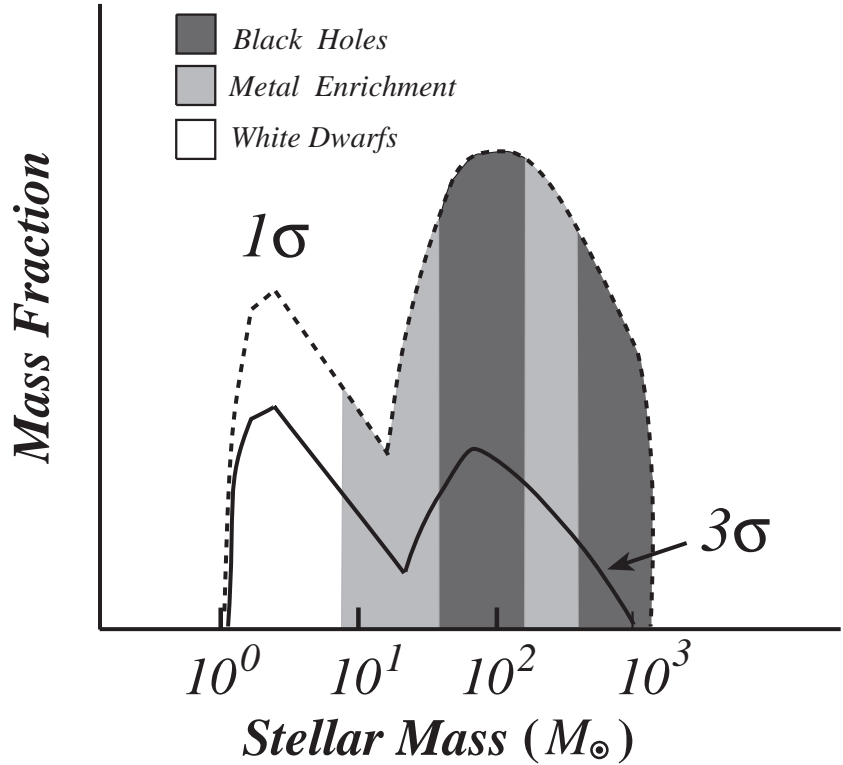}
\caption{Schematic initial mass function of Population III stars.
Solid and dashed lines are the IMFs for 3$\sigma$ and 1$\sigma$ density 
 fluctuations, respectively.
The IMF of Population III stars is likely to be bimodal and 
 approximated by a superposition of two 
 power-law like components with two different peaks of 
$m_{\rm p1}\approx 1-2 M_\odot$ and $m_{\rm p2}\approx {\rm 
 a \, few} \times 10-10^2 M_\odot$.
The relative height of the first peak descends
 with time compared to the second peak.
According to a recent theory of stellar evolution, 
stars with masses between $1-2M_\odot$ and $8M_\odot$
 are likely to evolve to white dwarfs that may reside
 in galactic halos as baryonic dark matter.
Stars with masses between $8M_\odot$ and $35M_\odot$
probably evolve to supernovae and eject heavy elements in 
 the intergalactic medium. 
Stars with masses between $35M_\odot$ and $10^2M_\odot$ 
 and greater than $250M_\odot$ are likely to collapse into 
 black holes and may be responsible for baryonic dark matter.
Stars with masses between $10^2M_\odot$ and $250M_\odot$ 
 probably explode as supernovae 
 and inject heavy elements into the intergalactic
 medium.
}
\label{fig:mass spectrum}
\end{figure}

\end{document}